\documentclass[10pt]{iopart}

\usepackage{iopams}
\usepackage{graphicx}
\usepackage{dcolumn}
\usepackage{bm}
\usepackage{multirow}
\usepackage{color}
\usepackage{hyperref}

\usepackage{dsfont}

\begin{document}
\newcommand{\hc}{H$_{c2}$}
\newcommand{\cso}{Cu$_{2}$OSeO$_{3}$}
\title[Magnonics with spin helices and skyrmions]{Collective spin excitations of helices and magnetic skyrmions: review and perspectives of magnonics in non-centrosymmetric magnets}
\author{Markus Garst$^1$, Johannes Waizner$^2$, and Dirk Grundler$^3$}
\address{$^1$ Institut f{\"u}r Theoretische Physik, TU Dresden, 01062 Dresden, Germany}
\address{$^2$ Institut f{\"u}r Theoretische Physik, Universit{\"a}t zu K{\"o}ln, Z{\"u}lpicher Str. 77a, 50937  K{\"o}ln, Germany}
\address{$^3$ Institute of Materials and Laboratory of Nanoscale Magnetic Materials and Magnonics,
School of Engineering, {\'E}cole Polytechnique F{\'e}d{\'e}rale de Lausanne, 1015 Lausanne, Switzerland}
\ead{\mailto{markus.garst@tu-dresden.de}, \mailto{dirk.grundler@epfl.ch}}
\vspace{10pt}
\begin{indented}
\item[]January 2017
\end{indented}

\begin{abstract}
Magnetic materials hosting correlated electrons play an important role for information technology and signal processing. The currently used ferro-, ferri- and antiferromagnetic materials provide microscopic moments (spins) that are mainly collinear. Recently more complex spin structures such as spin helices and cycloids have regained a lot of interest. The interest has been initiated by the discovery of the skyrmion lattice phase in non-centrosymmetric helical magnets. In this review we address how spin helices and skyrmion lattices enrich the microwave characteristics of magnetic materials. When discussing perspectives for microwave electronics and magnonics we focus particularly on insulating materials as they avoid eddy current losses, offer low spin-wave damping, and might allow for electric field control of collective spin excitations. Thereby, they further fuel the vision of magnonics operated at low energy consumption.
\end{abstract}

%
%
%
%
%

\section{Introduction: magnets for high-frequency applications}

\begin{figure}
\centering
    \includegraphics[width=0.5\textwidth]{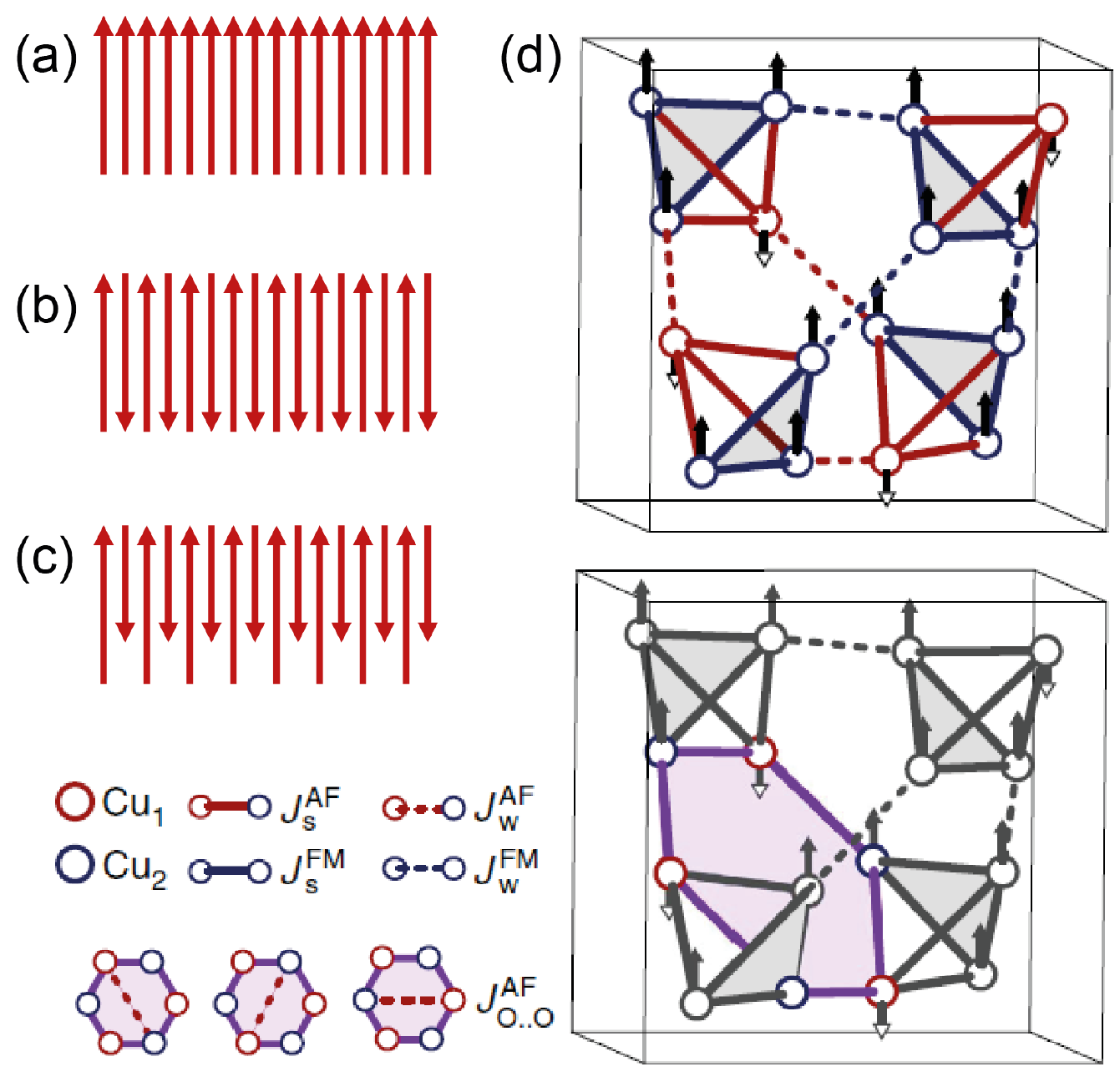}
	\caption{Spins configuration representing (a) ferromagnetic, (b) antiferromagnetic and (c) ferrimagnetic order assuming two sublattices of different spin moments. (d) Orientation of magnetic moments attributed to Cu ion sublattices in the chiral magnetic insulator \cso. Connecting lines categorize different exchange interactions: (top) first-neighbour ferromagnetic (FM) (blue) and antiferromagnetic (AF) (red) couplings. Strong (weak) bonds are indicated with solid (dashed) lines. Bottom: Further neighbour antiferromagnetic interaction realized on the diagonals of hexagons formed by alternating Cu$_1$ and Cu$_2$ sites. Compare legend on the left.
Figure \href{http://rdcu.be/oUJv}{Symmetrized TOF data along high-symmetry directions} rearranged from Ref. \cite{Portnichenko2016} published under CC-BY.}
	\label{fig1}
\end{figure}

Ferromagnetic materials played a dominant role in non-volatile data storage for decades. They dominated the perception of technology-relevant magnetic\footnote{Note that we address materials that show cooperative magnetic phenomena such as ferromagnetism, ferrimagnetism and antiferromagnetism. We do not consider diamagnetism and paramagnetism.} materials. In ferromagnets, microscopic magnetic moments (spins) align in parallel [Fig.~\ref{fig1} (a)] and lead to both a large static susceptibility $\chi$ and a large saturation magnetization $M_{\rm s}$ that allows one to encode data.

The functionality and relevance of antiferromagnets are often less appreciated as their spins are antiparallel [Fig.~\ref{fig1} (b)] and can completely compensate each other: $M_{\rm s}$ becomes zero for perfect magnetic order. In fact, when discussing antiferromagnets in his Nobel prize lecture Louis N{\'e}el argued in 1970 that antiferromagnetic materials were extremely interesting from the theoretical viewpoint, but did not seem to have any application \cite{Neel}. This viewpoint drastically changed after modern sensing applications and information technology required smaller and smaller ferromagnetic devices. In the course of miniaturization, they faced reduced stability due to environmental influences such as stray fields and increased operational temperatures. Nowadays the compensating spin structure of antiferromagnets plays an important role for stabilizing ferromagnets via exchange bias \cite{Meiklejohn1956}.

In Ref. \cite{Neel} Louis N{\'e}el discussed also the more complex spin structures of ferrites and ferrimagnets [Fig.~\ref{fig1} (c)] where sublattices of collinear spins but different $M_{{\rm s},i}$ (numbered with $i=1,2$) couple antiferromagnetically. The ferrites were technologically relevant since the 1930's. They are still produced in large amounts of several 100 000 t per year. Ferrimagnetic garnets are also of particular interest for applications. Following Ref. \cite{Neel} they are excellent insulators, can be prepared in large crystals, and can be used at very high frequencies in a large number of devices as their magnetization dynamics exhibits very sharp resonance lines. Accordingly, the insulating ferrimagnet yttrium iron garnet Y$_3$Fe$_5$O$_{12}$ (YIG) is exploited in microwave technologies such as circulators, band-pass filters and oscillators, ranging from radar applications to vector network analyzers \cite{Gurevich,Nikitov2015}. The relevant frequency regime ranges roughly from 1 to 100 GHz. Bulk YIG exhibits an extremely sharp resonance line and the smallest damping parameter $\alpha$ ever measured for magnetic resonances in ferro-, ferri- or antiferromagnetic materials. At room temperature its record value amounts to $\alpha=3\times10^{-5}$ \cite{Sparks1964}. Metals have not yet reached this low level of damping \cite{TomSilvaNeNatPhys}. Consequently, YIG has taken a key role in the research field of {\em magnonics} \cite{serga2010,Yu2014,Yu2016} where one aims at information processing using collective spin excitations in the form of spin waves (magnons) \cite{serga2010,2010KruglyakJPhysDApplPhys,2011Lenk,chumak2014,StampsRoadmap}. Spin waves as information carriers transport angular momentum but no charges. Thereby spin-wave based interconnects on chips \cite{ITRS}, spin-wave logic \cite{Khi2010,Khi2014} and on-chip microwave electronics avoid Joule heating and, in general, offer operation at low power consumption. Electric-field control of spin waves would offer even further reduced power consumption.

The advantages of magnonics can be harvested if one exploits spin-wave wavelengths smaller than 100 nm \cite{chumak2014}. Recently, nanopatterned magnetic materials with periodically modulated properties have provided a promising avenue in that magnonic grating couplers excited spin waves with wavelengths as small as 68 nm based on conventional microwave components operated at a few GHz \cite{Yu2016}. The wavelength of spin waves was smaller by a factor of about 300 000 compared to the corresponding electromagnetic wave in free space. This enormous wavelength reduction is a further key aspect of magnonics. The relevant magnonic grating couplers were realized by integrating periodic arrays of metallic ferromagnetic nanodisks to a thin YIG film \cite{Yu2016}. Even earlier it was shown that bicomponent periodic lattices formed magnonic crystals (MCs) providing artificially tailored spin-wave band structures consisting of allowed minibands and forbidden frequency gaps \cite{Wang2010,Tacchi2012}. A periodic lattice of nanostripes \cite{Wang2010} or nanodisks \cite{Tacchi2012} introduced a periodic potential which induced Bragg scattering of spin waves \cite{Krawczyk2013}. Periodically arranged domain walls have later been suggested as a further interesting building block for MCs. Here, non-collinear spin structures if realized in a plain film would avoid nanopatterning and allow for MCs with reconfigurable properties \cite{Duerr2012}.

\begin{figure}
	\includegraphics[width=\textwidth]{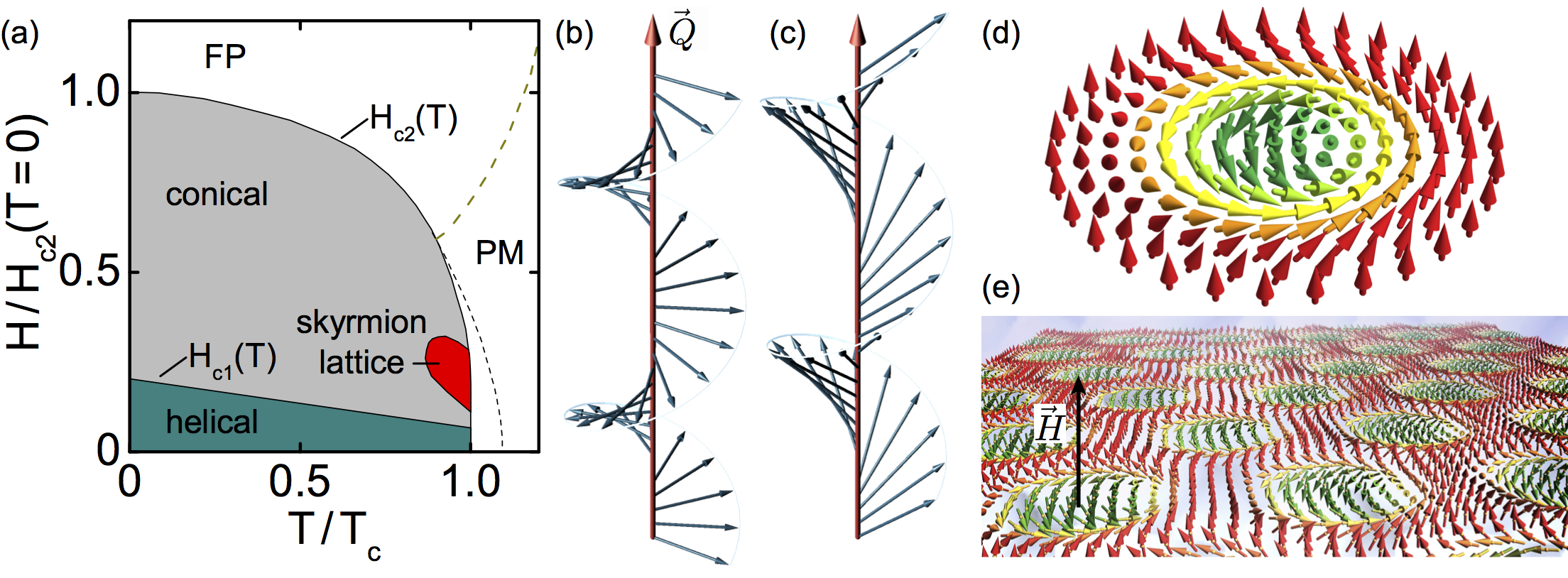}
	\caption{(a) Qualitative magnetic phase diagram of the cubic chiral helimagnets defining the critical field H$_{c2}$ and the critical temperature T$_c$ (PM: paramagnetic, FP: field polarized state) \cite{Schwarze}. Illustrations of (b) helical and (c) conical spin texture. $Q$ is the helical propagation vector. (d) Spin texture of an individual magnetic Skyrmion (Bloch-type). (e) Skyrmion lattice forming in a plane perpendicular to the applied field $H$.
}
	\label{fig2}
\end{figure}

Interestingly, chiral magnetic materials possess magnetic properties that are intrinsically periodic (Fig. \ref{fig2}).
In their case, non-collinear spin structures are stabilized by the Dzyaloshinsky-Moriya interaction (DMI) attributed to relativistic spin-orbit coupling \cite{1980BakJensen}. Examples are materials such as the metals MnSi, FeGe, the semiconductor Fe$_{0.8}$Co$_{0.2}$Si and the insulator Cu$_2$OSeO$_3$ \cite{2012SekiScience} that all possess a non-centrosymmetric cubic crystal structure with space group P2$_1$3 (Tab. \ref{Table1}).
\begin{table}
\caption{\label{MatOverview}
Overview and parameters of selected skyrmion-hosting materials.
}
\footnotesize
\begin{tabular}{@{}llllllll}
\br
Material & Space & Type & Pitch length & $T_c$ [K] & $\chi_{\rm con}^{\rm int}$ & magnetic
& inelastic\\
 &group& &$\frac{2\pi}{Q}$ [nm]& & &resonances &neutron\\
  & & & & & & &scattering\\
\mr
MnSi$^a$&P2$_1$3&metal&18&29& 0.34 &\cite{Date1977,Schwarze}&  \cite{2010Janoschek,Kugler2015}\\
\cso$^a$&P2$_1$3&insulator&60&58 & 0.65 &\cite{2012OnosePhysRevLett,Schwarze,2013OkamuraNatCommun} & \cite{Portnichenko2016,Tucker2016}\\
Fe$_{0.8}$Co$_{0.2}$Si$^a$&P2$_1$3&semiconductor&34&28&1.76&\cite{Schwarze,Watanabe1985}&\\
FeGe$^a$&P2$_1$3&metal&70&278&3.43&\cite{Zhang2016}$^c$&\cite{Siegfried2016}  \\
\mr
GaV$_4$S$_8$$^b$&R3m&semiconductor&22&11&&\cite{Ehlers2016a,Ehlers2016b}&\\
\br
\end{tabular}\\
$^{a}$helical and Bloch-type skyrmions; $^b$cycloidal and N{\'e}el-type skyrmions; $^c$GHz resonances were measured on FeGe that was deposited as a thin film on a substrate
\label{Table1}
\end{table}
\normalsize
They share the same magnetic phases that include periodic helical and conical spin structures as well as a skyrmion phase, see Fig.~\ref{fig2}. Skyrmions are particle-like spin textures (spin solitons) that exhibit nanometer-sized dimensions and a long lifetime due to topological protection \cite{Nagaosa13,Garst2016}.  They were first discovered in the magnetic phase diagram of the metallic chiral magnet MnSi \cite{2009MuhlbauerScience,Pfleiderer2010}. In bulk materials, they form a periodic magnetic skyrmion lattice (SkL) [Fig. \ref{fig2} (e)] with hexagonal symmetry, sometimes also called skyrmion crystal \cite{2010YuNature}.

Individual skyrmions are already foreseen to advance spintronics \cite{0022-3727-47-19-193001} and to provide a new platform for the race track memory \cite{2013FertNatureNano,2013SampaioNatNanotech,2013IwasakiNatureNano}, for magneto-logic \cite{2015ZhangSciRep} and GHz oscillators \cite{2015ZhangNJPhys,Kang2016,0022-3727-49-42-423001}. The term {\em skyrmionics} summarizes the research efforts that aim at skyrmion-based electronics. Here ultrathin metallic layers exhibiting spin-orbit coupling and interfacial DMI are currently preferred \cite{Heinze2011,Romming2013} as they provide magnetic skyrmions at room temperature \cite{Soum2016} and, at the same time, allow for the exploitation of existing thin-film deposition techniques and nanotechnology. Skyrmionics and related spintronics applications based on thin films with interfacial DMI were recently reviewed by G. Finocchio et al.~\cite{0022-3727-49-42-423001} and W. Kang et al.~\cite{Kang2016}.

\begin{figure}
\centering
	\includegraphics[width=8cm]{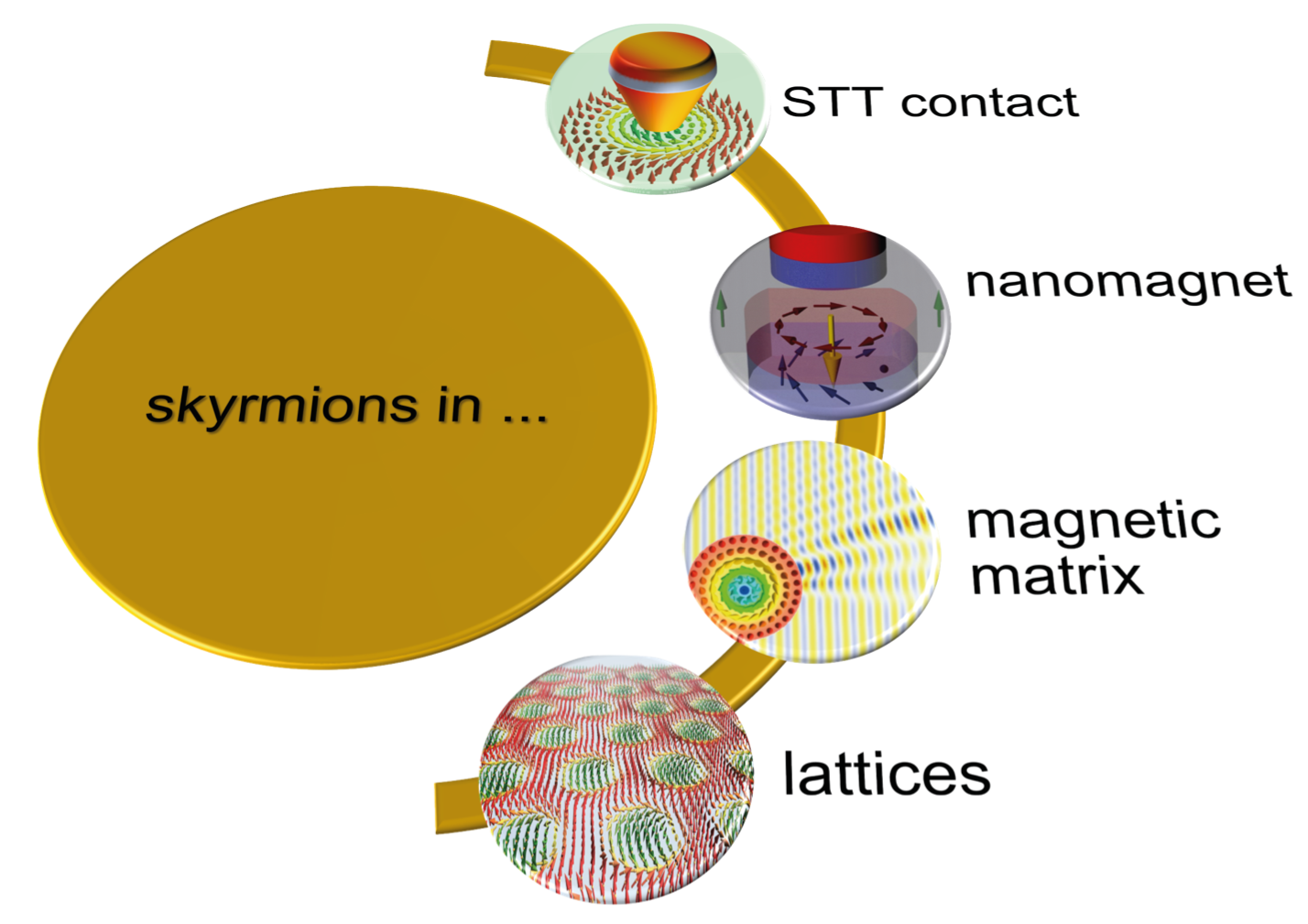}
	\caption{Four different scenarios in which skyrmions have been realized so far. All four scenarios have implications for magnonics, i.e., the control and manipulation of microwave signals by exploiting spin waves. Using spin-transfer torque (STT) a dynamically stabilized skyrmion can be obtained. Second figure from the top: \href{http://rdcu.be/oUK1}{Schematic diagram of the artificial skyrmion lattices and measured magnetic hysteresis loops} from Ref. \cite{Gilbert2015} published under CC-BY, rearranged from original.}
	\label{figIntro}
\end{figure}

For microwave- and magnonics-related applications also the initially discovered skyrmion lattice is intriguing (Fig. \ref{figIntro}).
However, the metallic layers and bulk metals are not suitable as the damping parameter $\alpha$ is large. Instead, insulators are needed. When aiming at high-power applications insulators in particularly bulk form are key. We thus consider the bulk material \cso\ with DMI to be the prototypical helimagnet when discussing {\em magnonics} with chiral magnets.

This paper is organized as follows. We first review spectroscopy data obtained in the GHz frequency regime on bulk materials hosting the skyrmion lattice phase \cite{KoralekPRL2006}. So far experiments have been performed at low temperatures as critical temperatures $T_{\rm c}$ of relevant materials such as MnSi and \cso\ are below room temperature. Spin wave spectroscopy on the alloy CoZnMn that hosts a skyrmion lattice at room temperature \cite{Tokunaga2015} has not yet been published. However, the spin dynamics of \cso\ have been studied extensively \cite{2010KobetsLTP,2010BelesiPhysRevB,2012SekiPhysRevB2,2012MaisuradzePPRL,2014LevacicPRB,Zhang2016,2012WhiteJPhysCondensMatter,2015OgawaSciRep,2015SaitohJAPL,2015Ruff}. Recently Mochizuki and Seki have thoroughly reviewed the spin dynamics of \cso\ focussing on excitation at the $\Gamma$ point, i.e., the uniform mode \cite{0953-8984-27-50-503001,2012MochizukiPhysRevLett,2012OnosePhysRevLett}. They discussed in detail magnetoelectric phenomena and the intriguing microwave diode effect occurring in bulk helimagnets
\cite{MochiSeki2013,2013OkamuraNatCommun,2015MochizukiPhysRevLett.114.197203,2015Okamura}. In this paper we review the full band structures of collective spin excitations ranging from zero to large wave vectors $k$ \cite{Kataoka1987,2010Janoschek} as they are key for applications in magnonics \cite{Stancil}. Our discussion addresses the prototypical insulating ferrimagnet \cso\ as, at this point it offers the smallest spin-wave damping \cite{Schwarze} and has shown the intriguing magnetochiral effects \cite{0953-8984-27-50-503001,2012MochizukiPhysRevLett,2012OnosePhysRevLett}. The properties of \cso\ particularly motivate the special attention of chiral magnets in magnonics. We then present theoretical considerations how to describe the dynamics in the complex spin structures. We compare theoretical aspects with experimental findings. In the final section we discuss prospects of spin helices and skyrmions in magnonics considering different scenarios (Fig. \ref{figIntro}).

\section{Experiments on GHz excitations in materials hosting the skyrmion lattice phase}

Spin dynamics in the chiral magnet MnSi was investigated early in the 1970's and 1980's \cite{Date1977,Ishikawa1985}, well before the discovery of the skyrmion lattice in the same material in 2009 \cite{2009MuhlbauerScience}. Since then, the interest in metallic MnSi and in its spin dynamics properties \cite{Schwarze} has largely increased. The skyrmions identified in MnSi represent so-called Bloch-type skyrmions as illustrated in Fig.~\ref{fig2}. In this case, the spins rotate in a {\em helical} manner between individual skyrmions. The plane of the skyrmion lattice is here perpendicular to the applied magnetic field. Later on similar skyrmions were found in other compounds of the same material class with the space group P2$_1$3 like \cso\,\cite{2012SekiScience,2012AdamsPhysRevLett,2012SekiPhysRevB}. In \cso\ the copper ions carry each a spin momentum $\frac{1}{2}$ and are arranged in Cu$_4$ tetrahedra as sketched in Fig. \ref{fig1} (d). In each tetrahedron the spins are ordered ferrimagnetically in that one Cu spin [red circles in Fig. \ref{fig1} (d)] is antiparallel with respect to the three other ones. In the field-polarized state each tetrahedron provides total spin ${\rm S}=1$ to the saturation magnetization $M_{\rm s}$ (compare Fig. 2 (d) in Ref. \cite{Janson2014}).

\begin{figure}
	\includegraphics[width=\textwidth]{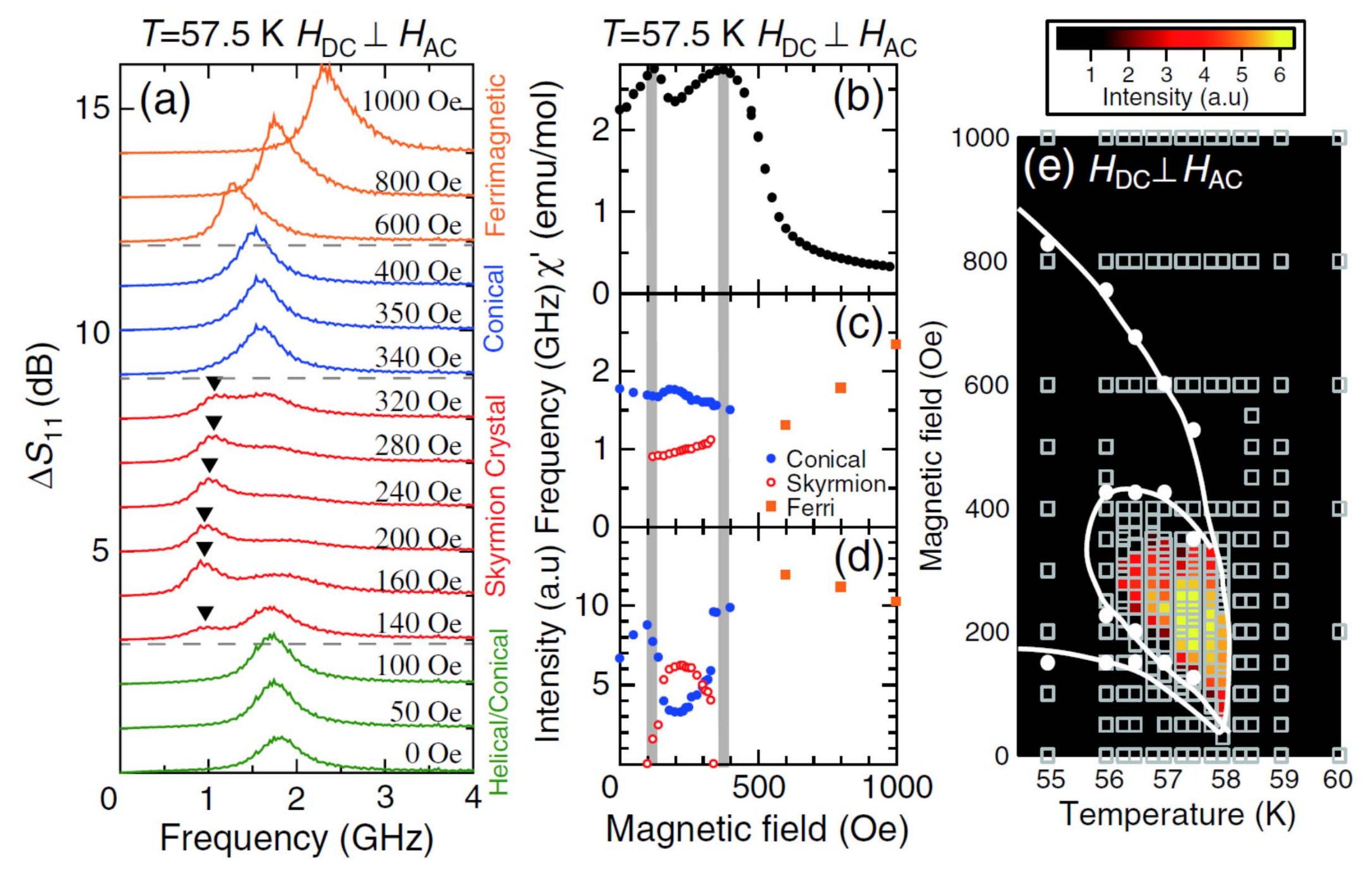}
	\caption{
	(a) The microwave absorption spectra $\Delta S_{11}$ for $H_{\rm DC}\perp H_{\rm AC}$ at various magnetic fields at 57.5 K. (b) Field dependence of the magnetic susceptibility. (c),(d) The (c) frequency and (d) intensity of the magnetic modes of (a). The closed and open circles and closed squares in (c),(d) correspond to the excitations in conical magnetic, Skyrmion
lattice, and field-polarized states, respectively. (e) The intensity of the counterclockwise (CCW) mode is only finite within the skyrmion lattice phase of the phase diagram. The intensity was determined for temperatures and fields as indicated by the squares.
Reprinted figure with permission from \href{https://doi.org/10.1103/PhysRevLett.109.037603}{Y. Onose et al., Phys. Rev. Lett. 109, 037603, 2012}. Copyright (2012)
by the American Physical Society.}
	\label{figOnose}
\end{figure}

Pioneering experiments on characteristic eigenmodes of the skyrmion lattice phase were performed on the Bloch-type skyrmions
by Onose et al.~exploring the ferrimagnetic insulator \cso\ \cite{2012OnosePhysRevLett}. Using a spectroscopy technique based on a broadband microwave transmission line Onose et al. detected three distinct modes in the few GHz frequency regime, see Fig. \ref{figOnose}.
The characteristic modes observed for the skyrmion lattice were consistent with an earlier theoretical prediction by Mochizuki \cite{2012MochizukiPhysRevLett} and termed clockwise (CW), counterclockwise (CCW) and breathing mode.
These modes obey different selection rules: whereas the CW and CCW mode are excited by an oscillating magnetic field $H_{\rm AC}$ located within the plane of the skyrmion lattice, the breathing mode couples to an ac field aligned perpendicular to the plane.
The employed technique allowed for the detection of one distinct mode in the field-polarized (FP) phase and two modes labelled as -Q and +Q in the helical (H) and conical (C) phases. In Ref.~\cite{2012OnosePhysRevLett} the selection rules and signal strengths of the different modes suggested an excitation via the magnetic field components of the microwaves.

Earlier published Raman and far-infrared spectroscopy spectra taken on \cso\ reported also several resonances \cite{Gnezdilov2010,Miller2010}. They resided at THz frequencies and were observed later also in electron spin resonance experiments performed at high frequencies \cite{Ozerov2014}.
An effective Heisenberg model identified such high-frequency modes as spin excitations in high-energy magnon bands of \cso\ \cite{PhysRevB.90.140404}. These are determined by the intratetrahedron exchange energy [Fig. \ref{fig1} (d)] that does not have a counterpart in e.g. the prototypical Skyrmion-hosting MnSi. The high-energy magnon bands are hence specific to \cso. At the same time, their THz frequencies are far beyond the frequencies that currently play a dominant role in the research field of magnonics and in modern information technology.
\begin{figure}
\centering
	\includegraphics[width=0.9\textwidth]{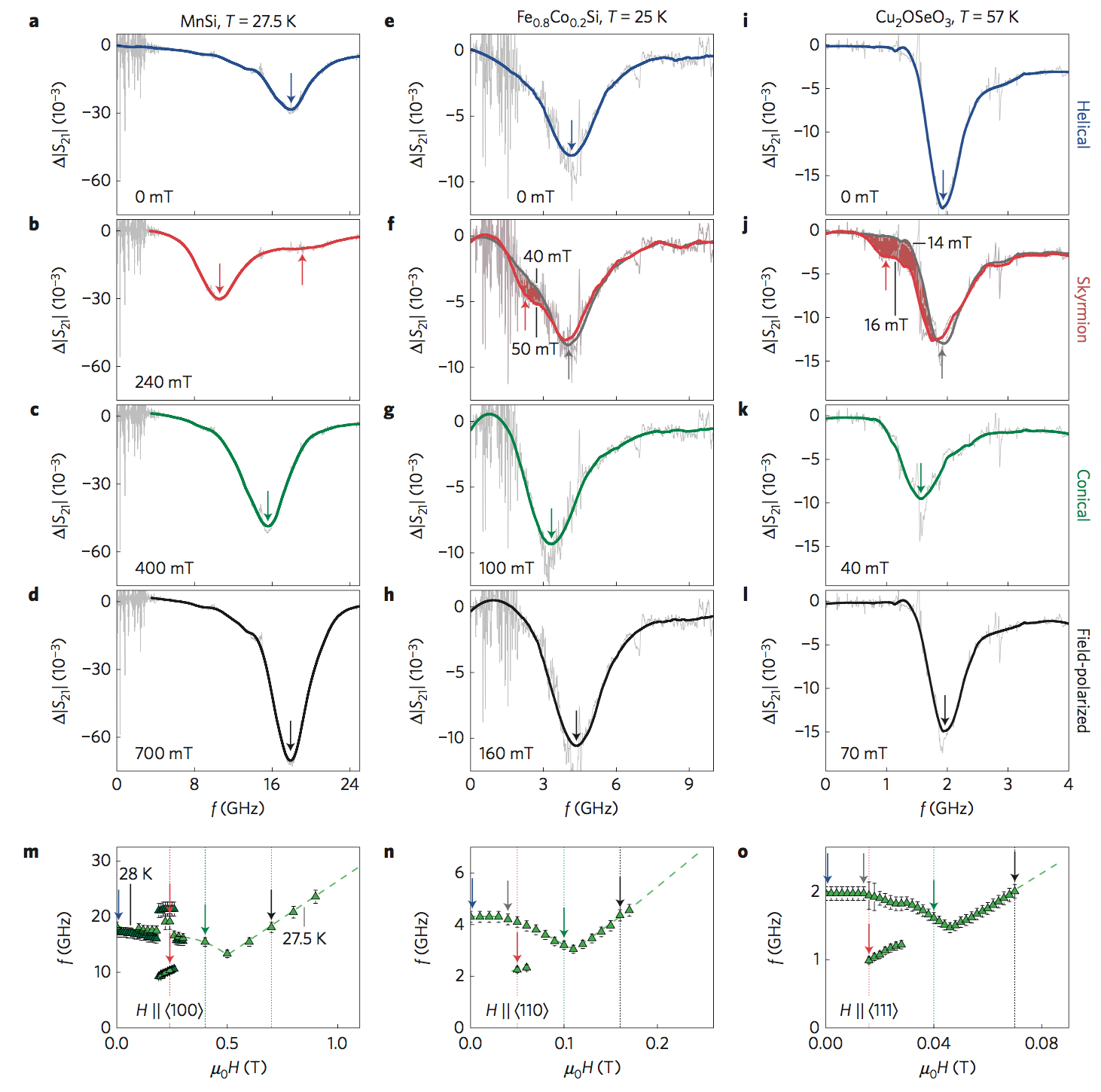}
	\caption{
	(a) to (l) Spectra obtained on MnSi, Fe$_{0.8}$Co$_{0.2}$Si and Cu$_2$OSeO$_3$ at various magnetic fields for temperatures just below the relevant $T_{\rm c}$. A smoothed curve was added as a guide to the eye.
(m) to (o) Corresponding resonance frequencies as a function of magnetic field. Note the different frequency scales. Characteristic resonance frequencies are marked by arrows. Dashed coloured lines indicate field values for which spectra are depicted in the upper rows. The shading between the curves in (f) and (j) highlights the occurrence of a low-frequency mode. Reprinted by permission from Macmillan Publishers Ltd, part of Springer Nature: Nature Materials (T. Schwarze {\em et al.}, Nat. Mater. 14, pages 478 -- 483), copyright (2015). Taken from Ref. \cite{Schwarze}. }
	\label{figSchwarze}
\end{figure}

In contrast, the low-energy spin excitations explored in the pioneering work by Onose et al.~are characteristic for chiral magnets of the P2$_1$3 material class and fall in the technologically relevant regime of GHz frequencies. In Ref. \cite{Schwarze} it was shown that the three modes detected in the SkL of \cso\ and the two modes -Q and +Q in the helical and conical states represent universal excitations of these phases. Schwarze {\em et al.} explicitly demonstrated that they are shared by the metal MnSi, the semiconductor Fe$_{0.8}$Co$_{0.2}$Si and the insulator \cso, see Fig.~\ref{figSchwarze}.

A member of a different class of skyrmion-hosting materials with a rhombohedral crystal structure and space group R3m \cite{Pocha2000} is GaV$_4$S$_8$. The presence of a magnetic skyrmion phase in bulk GaV$_4$S$_8$ was recently reported in Ref.~\cite{Kezsmarki2015a}. Here, the skyrmion lattice consists of N{\'e}el-type skyrmions where the spins rotate in a {\em cycloid} manner between individual skyrmions. The reduced crystal symmetry in this compound results in uniaxial magnetic anisotropies and favours the plane of the skyrmion lattice to be oriented along the crystallographic [111] direction. GaV$_4$S$_8$ shows a similar characteristic set of low-energy spin excitations as the
P2$_1$3 compounds. However, the relatively large magneto-crystalline anisotropy introduces a different hierarchy of resonance frequencies within the skyrmion lattice phase \cite{Ehlers2016a,Ehlers2016b}.

In the following section, we focus on chiral magnets with P2$_1$3 crystal symmetry and review the universal aspects of their low-energy collective spin excitations.
The universal characteristics allow one to functionalize the spin structures of chiral magnets over a broad frequency regime. In Fig. \ref{figSchwarze} frequencies range from about 1 to 30 GHz. At the same time, one can exploit magnetic insulators, semiconductors and metals, i.e., the three technologically relevant material classes. The broad range of available conductivities is particularly advantageous when combining magnonics with e.g. charge-based spintronics \cite{chumak2014}. We review the theoretical understanding and highlight aspects relevant for magnonics. We cover the energy and wavelength regime that is consistent with on-chip microwave technologies based on transmission lines and coplanar waveguides.

\section{Theoretical considerations about spin excitations in chiral magnets}

In the following we focus on the low-energy limit where the magnetization field can be treated in the continuum approximation. We neglect amplitude fluctuations and consider only the orientation of the magnetization, $\vec M(\vec r) = M_{\rm s} \hat n(\vec r)$, represented by the unit vector $\hat n$ and amplitude $M_s$.
Note that, e.g., MnSi and FeGe are itinerant chiral magnets where $M_{\rm s}$ possesses a residual magnetic field dependence. Moreover, Cu$_2$OSeO$_3$ is a ferrimagnet, for which $M_{\rm s}$ and $\hat n(\vec r)$ effectively represent the size and orientation of the ferrimagnetic moment, respectively. Intratetrahedron spin excitations involving the strong exchange interactions J$^{\rm FM}_{\rm s}$ and J$^{\rm AF}_{\rm s}$ [Fig. \ref{fig1} (d)] are not considered.
The magnetization field $\hat n$ is governed by the free energy functional $F = F_0 + F_{\rm dip} + F_{\rm aniso}$. In the cubic chiral magnets, the first part $F_0 = \int d\vec r \mathcal{F}_0$ possesses the density
\begin{equation} \label{F0}
\mathcal{F}_0 = A (\partial_\alpha \hat n_i)^2 + D \hat n ( \vec \nabla \times \hat n) - \mu_0 M_s \vec H \hat n
\end{equation}
where $\alpha, i = 1,2,3$ and summation over repeated indices is implied. $A$ is the exchange stiffness constant, $D$ denotes the strength of the Dzyaloshinskii-Moriya interaction, and $\vec H$ is the applied magnetic field. The sign of $D$ depends on the chirality of the atomic crystal structure; here, we assume $D>0$ resulting in right-handed chiral magnetic structures. For sufficiently small applied fields, the competition between the exchange interaction and the DMI stabilizes spatially modulated textures with a typical wavevector given by $Q = D/(2A)$ [Fig. \ref{fig2} (b) and (c)]. The DMI can also be expressed $\hat n ( \nabla \times \hat n) = \hat n^T (-i \vec {\bf L} \vec \nabla) \hat n$ in terms of the spin-1 operator $\vec {\bf L} = ({\bf L}^1, {\bf L}^2, {\bf L}^3)$ defined by the antisymmetric tensor
\begin{equation} \label{Spin1}
{\bf L}^\alpha_{ij} = - i \varepsilon_{ij\alpha},
\end{equation}
which is the generator of rotations. It fulfils the  spin-1 algebra $[{\bf L}^\alpha, {\bf L}^\beta] = i \varepsilon_{\alpha\beta \gamma} {\bf L}^\gamma$ and $\vec {\bf L}^2 = 2\, \mathds{1}$.

The second part comprises the dipolar interaction
\begin{equation} \label{Fdip}
F_{\rm dip} = \frac{\mu_0 M_{\rm s}^2}{2} \int d\vec r d{\vec r}{\,'} \hat n_i(\vec r) \chi_{{\rm dip},ij}^{-1}(\vec r - \vec r{\,'}) \hat n_j(\vec r{\, '}).
\end{equation}
The Fourier transform of the susceptibility $\chi_{{\rm dip},ij}^{-1}(\vec k)$ depends on the relative amplitude of the wavevector compared to the inverse linear size of the macroscopic sample $L$. For large wavevectors $|\vec k| \gg 1/L$, it is given by $\chi_{{\rm dip},ij}^{-1}(\vec k) =  \vec k_i \vec k_j/\vec k^2$. For small wavevectors $|\vec k| \ll 1/L$, it is determined by the demagnetization factor $N_{ij}$ of the sample $\chi_{{\rm dip},ij}^{-1}(\vec k) = N_{ij}$; for an ellipsoidal sample $N_{ij} = N_i \delta_{ij}$ with $N_x + N_y + N_z = 1$. A special situation arises for a highly anisotropic sample like a thin film which we will discuss below.

The third part $F_{\rm aniso}$ contains the magnetic and magnetocrystalline anisotropies. The former might be relevant for thin films whereas the latter is essential for the description of the helix reorientation transition at the first critical field $H_{c1}$ \cite{Bauer2016}. Here, for simplicity, we will not consider the modifications close to $H_{c1}$ [Fig. \ref{fig2} (a)] and neglect the magnetocrystalline anisotropies. We will include however the magnetic anisotropy in the discussion of thin films. In Tab. \ref{Table2} we summarize parameters that are relevant for the bulk materials and their free energy functional.

The equilibrium magnetization profile $\hat n_{\rm eq}(\vec r)$ is obtained by minimizing the free energy functional and solving the resulting Euler-Lagrange equations. This also includes certain boundary conditions that must be fulfilled at the surfaces of the sample. From the part in Eq.~(\ref{F0}), the following boundary conditions arise \cite{Rohart2013,Meynell2014}
\begin{equation} \label{BC}
\hat s \vec \nabla \hat n - Q \hat s \times \hat n \Big|_{\rm boundary} = 0
\end{equation}
where $\hat s$ is the unit vector orthogonal to the sample boundary. Note that the Dzyaloshinskii-Moriya interaction leads to an effective "pinning" of the spins at the boundary. We will not consider explicit boundary terms in the free energy functional which might modify Eq.~(\ref{BC}). The effect due to Eq.~(\ref{BC}) is negligible for bulk samples. However, in thin films it becomes important for the equilibrium magnetization and magnon spectrum as will be discussed in sect. \ref{sec:SpinWavesThinFilms}.

The DMI favors spatially modulated textures so that $\hat n_{\rm eq}(\vec r)$, in general, will depend on the spatial coordinate $\vec r$. In the linear spin-wave approximation, the magnon modes are obtained by expanding the energy functional around the equilibrium state. For this purpose, we introduce the local orthonormal frame $\hat e_i(\vec r) \hat e_j(\vec r) = \delta_{ij}$ with $\hat e_3 = \hat n_{\rm eq}$ and $\hat e_1 \times \hat e_2 = \hat e_3$. The spin excitations will be parametrized by the complex wavefunction $\psi$ in a standard fashion
\begin{eqnarray} \label{Expansion}
\hat n &= \hat e_3 \sqrt{1-2\frac{g \mu_B}{M_s} |\psi|^2} + \sqrt{\frac{g \mu_B}{M_s}}\Big( \psi \hat e_+ + \psi^* \hat e_-\Big)
\end{eqnarray}
where $\hat e_\pm = \frac{1}{\sqrt{2}} (\hat e_1 \pm i \hat e_2)$. The factor $\frac{g \mu_B}{M_s}$ is introduced so that the product $|\psi|^2$, representing the probability density of magnons, possesses the units of inverse volume. The equation of motion for the magnetization \cite{Gurevich} reads $\partial_t \hat n = -\gamma \hat n \times \vec B$ with $\gamma = g \mu_B/\hbar > 0$ and the effective magnetic field $\vec B = - \frac{1}{M_s} \frac{\delta F}{\delta \hat n}$. Expanding it up to linear order in the magnon wavefunction, we obtain the effective wave equation for the magnon modes that we discuss for various cases in the following. We show that in general the $U(1)$ symmetry associated with the phase of the $\psi$ wavefunction is broken. This reflects the fact that spin angular momentum carried by the magnons is not conserved because of spin-orbit coupling, dipolar interactions and the textured magnetization. For this reason, we introduce the spinor $\vec \Psi^T = (\psi, \psi^*)$ whose wave equation has the form
\begin{equation} \label{EVe}
i \hbar \tau^z \partial_t \vec \Psi(\vec r, t) = \mathcal{H} \vec \Psi(\vec r, t)
\end{equation}
where $\tau^z$ is a Pauli matrix and the Bogoliubov--deGennes Hamiltonian $\mathcal{H}$ is a $2 \times 2$ matrix operator. This Hamiltonian is constant in time so that we can limit ourselves to the solution of the stationary wave equation, $(\hbar \omega \tau^z - \mathcal{H})\vec \Psi(\vec r, \omega)=0$,  for the Fourier transform, $\vec \Psi(\vec r, t) = \int \frac{d\omega}{2\pi} e^{-i \omega t} \vec \Psi(\vec r, \omega)$. It is also convenient to introduce the retarded matrix Green's function
\begin{equation}
(\hbar (\omega+i0) \tau^z - \mathcal{H}) g(\vec r, \vec r^{\,'}; \omega) = \delta(\vec r-\vec r^{\,'}).
\end{equation}
The retarded dynamical magnetic susceptibility is then given by
\begin{equation} \label{Susc}
\chi_{ij}(\vec r, \vec r^{\,'};\omega) =
-g \mu_B \mu_0 M_s
(\hat e^+_i(\vec r) , \hat e^-_i(\vec r)) g(\vec r, \vec r^{\,'};\omega)
\left(\begin{array}{c}
\hat e^-_j(\vec r^{\, '}) \\ \hat e^+_j(\vec r^{\,'})
\end{array}
\right)
\end{equation}
The prefactor is chosen such that the projection of the spatial Fourier transform $\hat H_i \hat H_j\chi_{ij}(\vec k,\vec k';\omega)$ with $\hat H = \vec H/H$ reduces to the static dimensionless susceptibility $\chi = \partial M/\partial H$ for $\omega=0$ and $\vec k, \vec k' \to 0$. The imaginary part of Eq.~(\ref{Susc}) describes dissipation, and it is directly accessible in microwave absorption measurements as well as neutron scattering experiments.

\begin{table}
\caption{Experimental quantities in terms of the theoretical parameters $A$, $D$, and $M_s$ of Eq.~(\ref{F0}).
}
\centering
\footnotesize
\begin{tabular}{@{}lrl}
\br
pitch vector & $Q=$&$\frac{D}{2A}$\\[0.3em]
spin wave stiffness & $\mathcal{D} =$& $ \frac{2A g\mu_B}{M_s}$ \\[0.3em]
critical field energy & $g \mu_0 \mu_B H^{\rm int}_{c2} =$&
$\hbar \omega_{c2} =\mathcal{D}Q^2 = \frac{D^2 g\mu_B}{2A M_s}$ \\[0.3em]
susceptibility of the conical helix & $\chi^{\rm int}_{\rm con} =$&$\frac{M_s}{H^{\rm int}_{c2}} = \frac{2A\mu_0 M_s^2 }{ D^2}$ \\[0.3em]
\br
\end{tabular}
\label{Table2}
\end{table}

\subsection{Magnon excitations of the field-polarized state}
\label{subsec:field-polarized}

We revisit spin-wave excitations in the field-polarized regime of a large bulk sample [denoted by FP in Fig.~\ref{fig2}(a)]. A large magnetic field $H>H_{c2}$ is assumed to be applied along the $z$-axis, $\vec H = H \hat z$. The magnetization is polarized so that $\hat n_{\rm eq} = \hat z$. In this case, the local dreibein can be chosen to be independent of position, e.g., $\hat e_1 = \hat x$ and $\hat e_2 = \hat y$. The magnon Hamiltonian is then given by $\mathcal{H} = \mathcal{H}_0 + \mathcal{H}_{\rm dip}$ where $\mathcal{H}_0$ derives from Eq.~(\ref{F0}) and reads
\begin{equation}
\mathcal{H}_0 =
\mathcal{D}
(- \mathds{1} \vec \nabla^2 - i 2 Q \tau^z \hat z \vec \nabla)  + g \mu_B \mu_0 H \mathds{1}
\end{equation}
where we used the stiffness $\mathcal{D} = 2 A g \mu_B/M_s$ (Tab. \ref{Table2}).
In order to discuss the influence of the dipolar interaction, it is actually more convenient to switch from real to momentum space, $\mathcal{H}_0(- i \nabla) \to \mathcal{H}_0(\vec k)$. In momentum space, the Hamiltonian including the contribution from the dipolar interactions reads
\begin{equation} \label{PolBulk}
\mathcal{H} =
\mathcal{D}
(\mathds{1} \vec k^2 + 2 Q \tau^z \hat z \vec k)  + g \mu_B \mu_0 H_{\rm int} \mathds{1}
+
\frac{g \mu_B \mu_0 M_s}{2 k^2} \left(
\begin{array}{cc}
k_+ k_- &k_-^2 \\
k_+^2 & k_+ k_-
\end{array}
\right)
\end{equation}
where $H_{\rm int} = H - N_z M_s$ is the internal field, and we introduced $k_\pm = k_x \pm i k_y$.
The eigenfrequencies of the magnon spectrum then follow straightforwardly from solving the eigenvalue equation (\ref{EVe}) and are given by $\pm\omega(\pm\vec k)$ with
\begin{equation} \label{PolDisp}
\hspace{-2cm} \hbar \omega(\vec k) = 2 \mathcal{D} Q k_z + \sqrt{\left(\mathcal{D} k^2 + g \mu_B \mu_0 H_{\rm int}\right)\left(\mathcal{D} k^2 + g \mu_B \mu_0 H_{\rm int} + \frac{g \mu_B \mu_0 M_s k_\perp^2}{k^2}\right)}.
\end{equation}
where $\vec k_\perp = (k_x, k_y, 0)$ and $k = |\vec k|$.
The second term with the square root corresponds to the Herring-Kittel formula for a ferromagnet \cite{Herring:1951}. The DMI gives rise to the additional first term and leads to a shift of the dispersion along $k_z$, i.e., the direction in momentum space singled out by the magnetic field.

In particular, for vanishing perpendicular momentum, $\vec k_\perp = 0$, the dispersion reduces to a shifted parabola $\hbar \omega(k_z) = \mathcal{D} (k_z + Q)^2 + g \mu_B \mu_0 H_{\rm int} - \mathcal{D} Q^2$.
As a consequence, the magnon dispersion $\omega(k_z)$ is not symmetric with respect to $k_z$. This implies, for example, a finite group velocity $\partial_{k_z}\omega(k_z) \to 2 \mathcal{D} Q$ in the limit of small $k_z \to 0$. Corresponding nonreciprocal spin wave propagation was recently explored for Cu$_2$OSeO$_3$ \cite{Seki2016PRB}. The DMI-induced term in Eq. (\ref{PolDisp}) also results in an asymmetry in the dynamic susceptibility with respect to $k_z$. Introducing the spin-flip contributions for the dynamic susceptibility of Eq.~(\ref{Susc}), $\chi_{\pm\mp}(k_z,\omega) = \hat e^\mp_i \chi_{ij}(k_z,\omega) \hat e^\pm_j$, we obtain for its imaginary parts at $k_\perp = 0$
\begin{equation}
\chi''_{\pm\mp}(k_z,\omega) =\pm
\frac{\pi g \mu_B \mu_0 M_s}{\hbar} \delta(\omega \mp \omega(\pm k_z)).
\end{equation}
A magnon at a given momentum $k_z$ and energy $\omega(k_z)$ can be absorbed, but, different from a conventional ferromagnet, it cannot be emitted with the same energy as $\omega(-k_z) \neq \omega(k_z)$. This asymmetry in absorption and emission of magnons has been very recently observed in inelastic neutron scattering on MnSi \cite{Sato2016}.

The energy dispersion $\omega(\vec k)$, Eq.~(\ref{PolDisp}), is mimimal for $\vec k = Q \hat z$ with an energy gap $g \mu_B \mu_0 H_{\rm int} - \mathcal{D} Q^2$. This gap vanishes at the second critical field $H^{\rm int}_{c2}$ where the transition occurs between the field-polarized (FP) and the conical phase, see Fig.~\ref{fig2}(a). The stiffness $\mathcal{D} = g \mu_B \mu_0 H^{\rm int}_{c2}/Q^2$ is thus determined by the size of the pitch vector $Q$ and the value of $H^{\rm int}_{c2}$, which are both accessible from independent measurements \cite{Kugler2015,Grigoriev:2015,Siegfried2016}.

We comment on the  eigenmodes at zero wavevector attributed to uniform oscillations of the magnetization. In this limit, the Hamiltonian simplifies to
\begin{equation}
\hspace{-2cm} \mathcal{H}_{\rm uniform} =  g \mu_B \mu_0 H \mathds{1} + g \mu_B \mu_0 M_s \Big(- N_z \mathds{1} + \frac{N_x + N_y}{2} \mathds{1} + \frac{N_x - N_y}{2} \tau^x \Big).
\end{equation}
For the field-polarized phase, the eigenfrequencies
have the form of the well-known Kittel mode \cite{Kittel1948,Gurevich}
\begin{equation} \label{Kittel}
\hbar \omega = g \mu_B \mu_0 \sqrt{(H + (N_x  - N_z)M_s) (H + (N_y  - N_z)M_s)}.
\end{equation}
The uniform mode within a sample of large size is thus not affected by the DMI.

\subsection{Magnon excitations of the helical spin structure}

When the magnetic field $H$ is lowered below $H_{c2}$, the field-polarized state becomes unstable with respect to the formation of a conical spin helix. It is characterized by a helix axis that is aligned with the magnetic field.
The conical helix is realized in the chiral magnets within the field range between $H_{c1}$ and $H_{c2}$. If the field is lowered below $H_{c1}$ the helix might reorient its axis from the field direction into a crystallographic high-symmetry direction, either $\langle 111\rangle$ or $\langle 100\rangle$ depending on the material \cite{Bauer2016}.  This reorientation has its origin in the magnetocrystalline anisotropies $F_{\rm aniso}$. If the field is applied along the favoured crystallographic direction then $H_{c1} = 0$ in case of high-field cooling. In the following, we neglect these anisotropies and the associated complexities arising from the reorientation transition close to $H_{c1}$. We limit ourselves to the spin waves in the presence of a conical helix configuration. Here, we consider the situation of a bulk sample. Thin films will be discussed in section \ref{sec:SpinWavesThinFilms}.

\subsubsection{Spin-waves of the magnetic helix in bulk samples: finite wavevectors}
The conical helix configuration for a magnetic field applied along the $z$-axis is given by
\begin{equation} \label{ConHelix}
\hat n_{\rm helix}(z) = \sin \theta e^{-i Q z {\bf L}^z} \hat x+ \cos\theta \hat z =
\left(\begin{array}{c}
\sin \theta \cos (Q z) \\
\sin \theta \sin (Q z) \\
\cos \theta
\end{array} \right)
\end{equation}
It consists of a homogeneous part pointing along the magnetic field direction and a helical part that rotates as a function of the $z$-coordinate within the plane perpendicular to the field. The helical part is favored by the DMI and can be generated with the help of the $z$-component of the spin-1 operator $\vec {\bf L}$ introduced in Eq.~(\ref{Spin1}).

Plugging this Ansatz into the free energy functional and minimizing with respect to the cone angle $\theta$ one finds $M_s \cos \theta = \chi_{\rm con} H = \chi^{\rm int}_{\rm con} H_{\rm int}
$ with the susceptibility $\chi^{-1}_{\rm con} = (\chi^{\rm int}_{\rm con})^{-1} + N_z$. We assume here that the $z$-axis coincides with a principal axis of the sample characterized by the demagnetization factor $N_z$. The internal susceptibility is given by $\chi^{\rm int}_{\rm con} =  \mu_0 M_s^2/(2A Q^2)$. On the mean-field level, the differential susceptibility $\partial M /\partial H = \chi_{\rm con}$ is thus expected to be constant within the conical phase. This is  indeed  observed approximately in chiral magnets except for relatively high temperatures close to $T_c$ \cite{Bauer2010}. The experimentally determined value of the susceptibility $\chi^{\rm int}_{\rm con}$, see Table \ref{Table1}, provides an important combination of parameters that will enter the spin-wave spectrum. Moreover, at the second critical field $H_{c2}$ the angle $\theta = 0$ so that $\chi^{\rm int}_{\rm con} H^{\rm int}_{c2} = M_s$ which is consistent with the result of section \ref{subsec:field-polarized}.

The helix possesses the continuous screw symmetry. An arbitrary translation of the helix along the $z$-axis can be compensated by a rotation around the same axis leaving the helix invariant. Mathematically, this is reflected in the fact that $({\bf P}_z - Q {\bf L}^z) \hat n_{\rm helix}(z) = 0$ where ${\bf P}_z = - i \mathds{1} \partial_z$. An additional discrete symmetry arises in zero field where $\cos \theta = 0$. In this case, the helix is also invariant with respect to a $\pi$ rotation
of real and spin space around the $x$-axis. These symmetries will be reflected in specific properties of the spin wave excitations as discussed below.

In the conical helix phase, the magnon Hamiltonian deriving from the exchange free energy density $\mathcal{F}_0$ of Eq.~(\ref{F0}) reads \cite{Petrova2011,Kugler2015}
\begin{equation} \label{H0Helix}
\mathcal{H}_0 = \mathcal{D} [- \mathds{1} \vec \nabla^2  - i 2 \tau^z Q \hat n_\perp(z) \vec \nabla + \frac{Q^2 \sin^2\theta}{2} (\mathds{1} - \tau^x)].
\end{equation}
The magnetic helix gives rise to a periodic potential, $\hat n_\perp(z) = (\sin \theta \cos (Q z),\\ \sin \theta \sin (Q z),0 )$, for the magnons along the $z$-direction. The periodic potential leads to Bragg scattering that opens gaps in the energy spectrum and results in magnon bands with different index $n = 0,1,2,3,...$. In accordance with Bloch's theorem, the spin wave spectrum $\omega_n(\vec k)$ is periodic $\omega_n(\vec k + m \vec Q) = \omega_n(\vec k)$ for any $m \in \mathds{Z}$ with $\vec Q = Q \hat z$. In real space, relevant periodicities $2\pi/Q$ are on the order of 20 to 70 nm for the materials listed in Tab. \ref{Table1}.

The wave equation is in general not diagonal in momentum space. The full stationary equation including the dipolar interactions reads
\begin{eqnarray} \label{WEHelix}
- g^{-1}_0(\omega,\vec k) \vec \Psi_\omega(\vec k)
+ V(\vec k) \vec \Psi_\omega(\vec k - \vec Q) + V^*(\vec k) \vec \Psi_\omega(\vec k + \vec Q)
\\\nonumber
+ \sum_{ \alpha,\beta = -1,0,1, \atop \alpha \neq \beta} \mathcal{H}^{\alpha\beta}_{\rm dip}(\vec k - \alpha \vec Q)  \Psi_\omega(\vec k - (\alpha - \beta) \vec Q) = 0.
\end{eqnarray}
The periodic potential of Eq.~(\ref{H0Helix}) leads to the off-diagonal potential $V(\vec k) = \mathcal{D} Q  k_- \sin \theta \tau^z$ with $k_\pm = k_x \pm i  k_y$.
The diagonal part is summarized in the Green function
\begin{equation}
\hspace{-2cm} - g^{-1}_0(\omega,\vec k) = - \hbar \omega \tau^z + \mathcal{D} [\mathds{1} \vec k^2 + \frac{Q^2 \sin^2\theta}{2} (\mathds{1} - \tau^x)] + \sum_{\alpha = -1,0,1} \mathcal{H}^{\alpha\alpha}_{\rm dip}(\vec k - \alpha \vec Q)
\end{equation}
The contribution of the dipolar interaction (\ref{Fdip}) leads to the terms $\mathcal{H}^{\alpha\beta}_{\rm dip}$ whose explicit form is given in \ref{appendix}.
\begin{figure}
\centering
\includegraphics[width=0.9\textwidth]{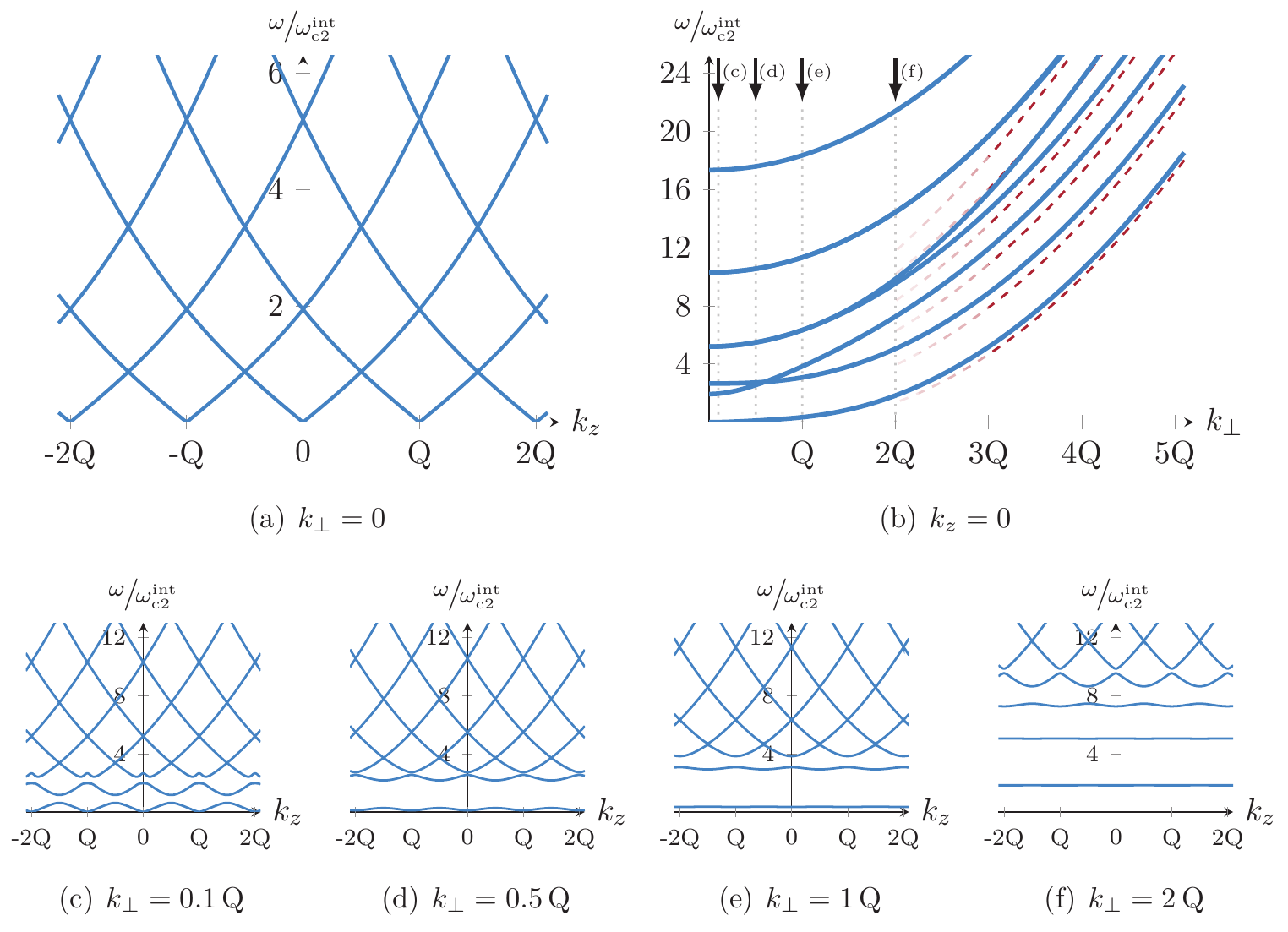}
\caption{Magnon spectrum in the presence of a helix with cone angle $\theta = \pi/2$ corresponding to zero magnetic field with $\chi_{\rm con}^{\rm int}=1.76$ corresponding to Cu$_{2}$OSeO$_{3}$. Bragg scattering off the helix gives rise to a magnon band structure as a function of wavevector $k_z$ along the helix axis. Spectrum for (a) $\vec k_\perp = 0$ of Eq.~(\ref{SpectrumHelix}) in the extended zone scheme and (b) $k_z = 0$. The red dashed lines corresponds to the asymptotics of Eq.~(\ref{HelimagnonSpectrumAsymptotic}) Arrows indicate cuts shown in the next panels. Panel (c) - (f) illustrate the opening of band gaps for increasing values of $k_\perp= |\vec k_\perp|$ and the emergence of flat dispersionless magnon bands. In panel (c) the little bump  of the third band  close to zone center is attributed to dipolar interactions.}
\label{Fig:HelixSpectrum}
\end{figure}
If the magnon only possesses a momentum along the $z$-direction, i.e., the direction of the helix, $\vec k = k_z \hat z$, the problem simplifies considerably. In this case, magnon gaps do not exist in the spectrum because the Fourier transform of the periodic potential, $V(\vec k)$, vanishes. One is left with the empty lattice model for spin waves \cite{Krawczyk2013}. The vanishing of Bragg scattering in this limit is attributed to the continuous screw symmetry of the helix. The spectrum can then be obtained exactly and reads
\begin{equation} \label{SpectrumHelix}
\hbar \omega(k_z)
= \mathcal{D} |k_z| \sqrt{k_z^2 + (1+\chi_{\rm con}^{\rm int}) Q^2 \left(1 - \left(\frac{H_{\rm int}}{H^{\rm int}_{c2}}\right)^2\right)}
\end{equation}
 in the extended zone scheme where the term $\chi_{\rm con}^{\rm int}$ arises due to the dipolar interaction, and $H_{\rm int}/H^{\rm int}_{c2} = H/H_{c2}$.

In the other limit of large perpendicular momenta $|\vec k_\perp| \gg Q$ with $\vec k_\perp\hat z=0$, the dipolar interaction and the last term in Eq.~(\ref{H0Helix}) can be effectively neglected. The wave equation for the partial Fourier transform $\vec \Psi_{\vec k_\perp,\omega}(z)$ then reduces to
\begin{equation}
\hbar \omega \tau^z \vec \Psi_{\vec k_\perp,\omega}(z) \approx \mathcal{D}[\mathds{1}(\vec k_\perp^2 - \partial_z^2) - i 2 \tau^z Q |\vec k_\perp| \sin \theta \cos(Q z - \alpha)]
\vec \Psi_{\vec k_\perp,\omega}(z)
\end{equation}
with $\vec k = |\vec k_\perp| (\cos \alpha, \sin \alpha , 0)$. It describes a particle with quadratic dispersion in a periodic cosine potential. This wave equation can be identified with the {\it Mathieu equation}. Interestingly, the strength of the periodic potential can be tuned by the size of the perpendicular momentum $\vec k_\perp$ allowing to tune the band structure from the weak-binding
to the tight-binding limit. For very large $|\vec k_\perp| \gg (n+1)^2 Q/\sin\theta$, the Bragg scattering is so strong that the band with index $n$ is basically flat, i.e., non-dispersive, and the spin waves are localized within the $z$-direction. In this limit, the periodic cosine potential $\cos(Q z - \alpha)$ can be expanded around its minima. Solving the resulting wave equation one obtains \cite{Janoschek:2010,Kugler2015}
\begin{equation} \label{HelimagnonSpectrumAsymptotic}
\hspace{-2cm} \hbar \omega_{n}(\vec k) \approx \mathcal{D}\Big[ \vec k_\perp^2 - 2 |\vec k_\perp| Q \sin \theta + 2 Q\sqrt{Q |\vec k_\perp| \sin\theta} \Big(n + \frac{1}{2}\Big) - \frac{Q^2}{8} (n^2 + n)\Big]
\end{equation}
where $\sin \theta = \sqrt{1-(H/H_{c2})^2}$. The third term  corresponds to the harmonic oscillator spectrum deriving from oscillations around the minima of the cosine potential. The last term is attributed to the anharmonicity of the potential. The spin wave spectra numerically solved for various values of $\vec k_\perp$ are shown in Fig.~\ref{Fig:HelixSpectrum}.

At the Brillouin zone center close to zero wavevector the energy of the lowest band vanishes. This low-energy spin wave is a Goldstone mode and protected by translational symmetry. Both, at zero magnetic field and in the absence of magnetocrystalline anisotropies $F_{\rm aniso}$ that would break the rotational invariance of the theory, the dispersion of this mode is particularly soft \cite{Belitz:2006,Radzihovsky2011},
\begin{equation} \label{GoldstoneMode}
\omega^2 \sim B k_z^2 + K k_\perp^4.
\end{equation}
It has the Landau-Peierls form characteristic for lamellar structures where $B$ and $K$ are the elastic constants of compression and splay \cite{ChaikinLubensky}, respectively. Such a mode would lead to a Landau-Peierls instability destroying long-range magnetic order of the conical helix. However, either a magnetic field or small magnetocrystalline anisotropies eventually lead to a $k_\perp^2$ contribution to the right-hand side of Eq.~(\ref{GoldstoneMode}) that stabilizes the helimagnetic order \cite{Belitz:2006}.

\begin{figure}
\centering
\includegraphics[width=0.5\textwidth]{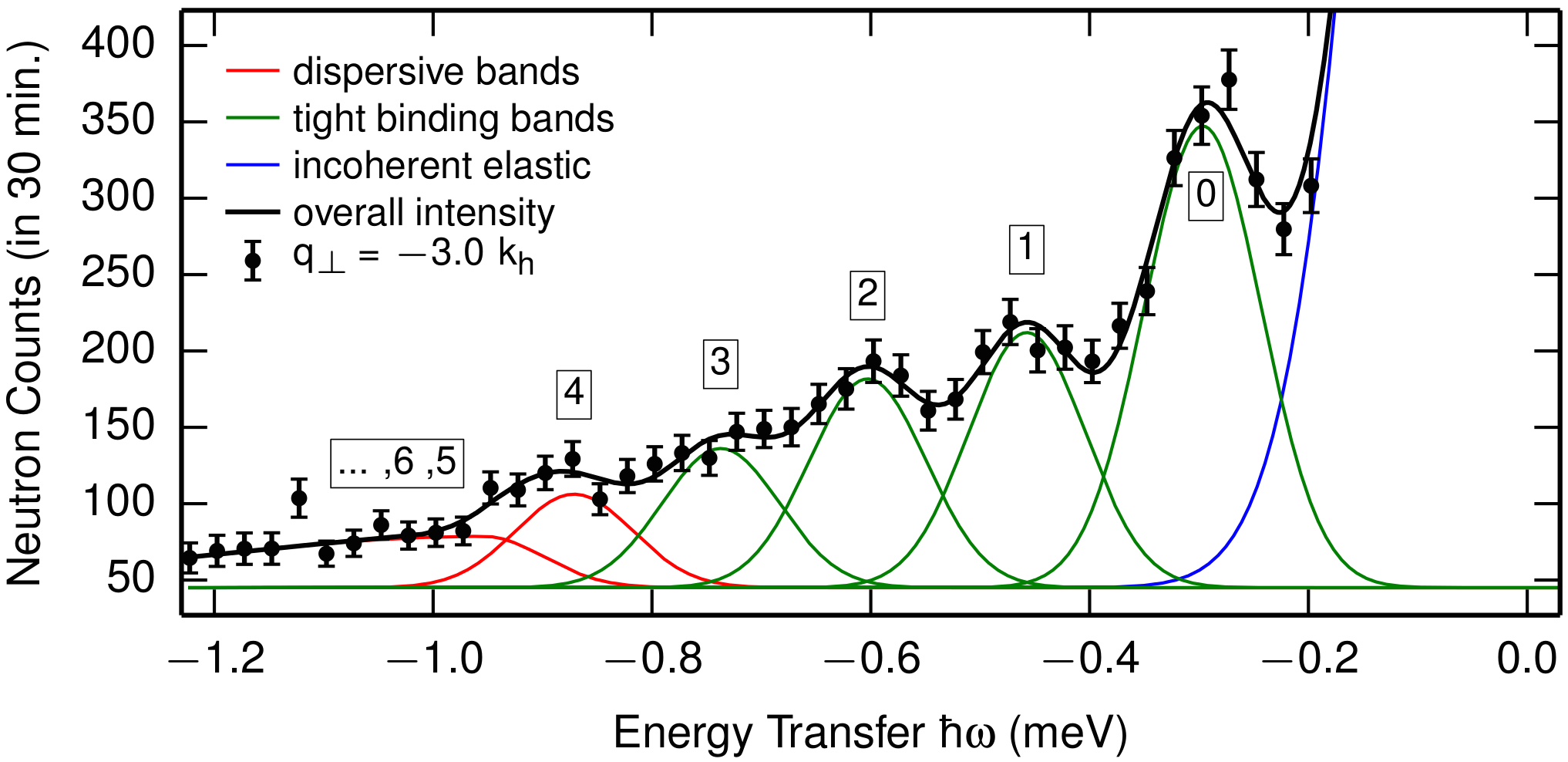}
\caption{Inelastic neutron scattering on MnSi \cite{Kugler2015}. Five bands are resolved as a function of energy loss by absorption of helimagnons at $|\vec k_\perp| = 3 Q$ and $k_z = 0$.}
\label{Fig:Kugler}
\end{figure}

The helimagnon spectrum has been measured in MnSi by inelastic neutron scattering \cite{Janoschek:2010,Kugler2015}. Kugler {\it et al.} \cite{Kugler2015} resolved five helimagnon bands as a function of energy as shown in Fig.~\ref{Fig:Kugler}. Such measurements have been also performed on Cu$_{2}$OSeO$_{3}$ \cite{Portnichenko2016,Tucker2016} but here the band structure could not be resolved because the associated energy scale $\mathcal{D} Q^2$ is an order of magnitude smaller than in MnSi.

\subsubsection{Dynamics of the spin helix at zero wavevector: uniform mode}

According to Eq.~(\ref{ConHelix}), the uniform magnetization is given by $\vec M = \hat z M_s \cos\theta = \hat z \chi_{\rm con} H$. The excitation of the uniform mode results in a
dynamic correction to $\vec M$ that, in first order in the magnon wavefunction, reads
\begin{eqnarray}
\delta \vec M(t) =
\frac{M_s}{V}\int d\vec r
\sqrt{\frac{g \mu_B}{M_s}} \frac{1}{2}{\rm Re}\Big\{(\hat e^+(z),  \hat e^-(z)) \vec \Psi(\vec r, t)
\Big\}
\end{eqnarray}
with the volume $V$. We can perform the real space integral using the representation
$\hat e_1 = (-\sin(Q z), \cos(Q z),0)$ and $\hat e_2 = (-\cos\theta \cos(Qz), - \cos\theta \sin (Qz),\sin \theta)$ for the vectors $\hat e^\pm = \frac{1}{\sqrt{2}}(\hat e_1 \pm i \hat e_2)$.

Consider first the $z$-component
\begin{equation}
\delta \vec M_z(t) =
\frac{\sqrt{g \mu_B M_s}}{2\sqrt{2}V}
\sin\theta\,  {\rm Re}\Big\{  i  (1,-1) \vec \Psi(0, t)\Big\}
\end{equation}
where $\vec \Psi(0,t) = \int d \vec r \, \vec \Psi(\vec r,t)$. It is determined by the magnon wavefunction at zero momentum $\vec \Psi(0, t)$, which coincides with the Goldstone mode of the helix associated with the breaking of translational symmetry.
Its eigenvector is therefore related to the derivative $\partial_z \hat n_{\rm helix}(z) \propto \hat e_1$. Comparing this with Eq.~(\ref{Expansion}) identifies the eigenmode $\vec \Psi^T(0, \omega) = \frac{1}{\sqrt{2}}(1,1)$ with eigenfrequency $\omega = 0$. The scalar product $(1,-1) \vec \Psi(0, t)$ determining $\delta \vec M_z$ thus exactly vanishes. So we conclude that  in linear order in the spin wave amplitude the mean magnetization does not vary in time along the $z$-direction, $\delta \vec M_z(t) = 0$, within the linear spin-wave approximation considered here.

\subsection*{Resonance frequencies of spin helix modes}

The uniform oscillation of the magnetization is therefore confined to the plane in spin space that is orthogonal to the magnetic field,
\begin{eqnarray} \label{HelixUniformModeXY}
\delta \vec M_x(t) = \frac{\sqrt{g \mu_B M_s}}{4\sqrt{2}V}
\times \\\nonumber\quad
{\rm Re}\Big\{i
\Big((-1-\cos\theta, -1+\cos\theta),(1-\cos\theta, 1+\cos\theta)\Big)
 \left(\begin{array}{c} \vec \Psi(\vec Q, t) \\ \vec \Psi(-\vec Q, t)\end{array}\right)
\Big\}
\\
\delta \vec M_y(t) = \frac{\sqrt{g \mu_B M_s}}{4\sqrt{2}V}
\times \\\nonumber\quad
 {\rm Re}\Big\{
\Big((1+\cos\theta, 1-\cos\theta),(1-\cos\theta, 1+\cos\theta)\Big)
 \left(\begin{array}{c} \vec \Psi(\vec Q, t) \\ \vec \Psi(-\vec Q, t)\end{array}\right)
\Big\}
\end{eqnarray}
where $\vec \Psi(\pm \vec Q,t) = \int d \vec r e^{\mp i \vec Q \vec r} \vec \Psi(\vec r,t)$.
The oscillations within this $(x,y)$ plane are determined by the two components $\vec \Psi(\pm \vec Q,\omega)$ of the magnon wavefunction. It turns out that these two components are decoupled from the other modes and governed by the wave equation
\begin{equation}\label{UniformModeHelix}
\left(\begin{array}{cc}
- g^{-1}_0(\omega,\vec Q) & \mathcal{H}^{1,-1}_{\rm dip}(0) \\
\mathcal{H}^{-1,1}_{\rm dip}(0) & -g^{-1}_0(\omega,-\vec Q)
\end{array}\right)
\left(\begin{array}{c}
\vec \Psi(\vec Q,\omega) \\
\vec \Psi(-\vec Q,\omega)
\end{array}\right) = {\bf 0}
\end{equation}
where the Green function reads explicitly
\begin{eqnarray}
-g^{-1}_0(\omega,\pm\vec Q) = -\hbar \omega \tau^z
+ \mathcal{D} Q^2 \Big[\mathds{1} + \frac{\sin^2\theta}{2} (\mathds{1} - \tau^x)]
\\ \nonumber
+\chi^{\rm int}_{\rm con} \left(\frac{\sin^2\theta}{2} (\mathds{1} - \tau^x)
+ \frac{N_x + N_y}{4}((1+\cos^2 \theta) \mathds{1} + \sin^2 \theta \tau^x \pm \cos\theta \tau^z)
\right)\Big].
\end{eqnarray}
The term in the last line is attributed to the dipolar interaction and, again, its strength is conveniently parametrized by $\chi^{\rm int}_{\rm con}$; note that $\mathcal{D} Q^2 \chi^{\rm int}_{\rm con}
= g \mu_B \mu_0 M_s$ (Tab. \ref{Table2}). The off-diagonal parts read
\begin{equation}
\mathcal{H}^{1,-1}_{\rm dip}(0) = \mathcal{D} Q^2 \chi^{\rm int}_{\rm con} \frac{N_x - N_y}{4}\Big(-\sin^2 \theta \mathds{1} - i 2 \cos \theta \tau^y - (1+\cos^2 \theta) \tau^x\Big)
\end{equation}
and $\mathcal{H}^{-1,1}_{\rm dip}(0) = (\mathcal{H}^{1,-1}_{\rm dip}(0))^\dagger$.
The solution of the $4\times 4$ matrix (\ref{UniformModeHelix}) determines the
eigenfrequencies and eigenvectors of the uniform oscillations of the magnetization within the conical helix phase.
\begin{figure}
\centering
\includegraphics[width=0.9\textwidth]{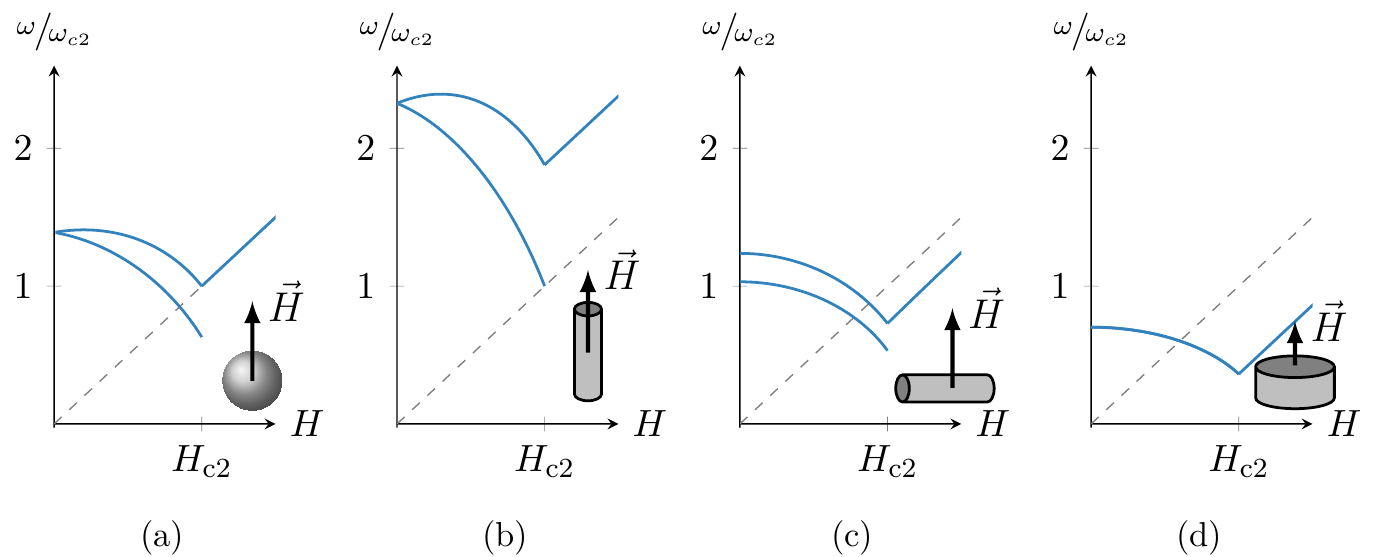}
\caption{Resonance frequencies of uniform modes in a chiral magnet with $\chi_{\rm con}^{\rm int} = 1.76$ corresponding to the value of Cu$_2$OSeO$_3$ normalized with respect to $\hbar \omega_{c2} = g \mu_B \mu_0 H_{c2}$.
The dc magnetic field $\vec H$ is applied along the $z$-axis and tunes between the conical and field-polarized phase with a phase transition at $H_{c2}$. There are two resonances $\pm Q$ in the conical phase, $H<H_{c2}$, whose frequencies and field dependencies strongly depend on the sample shape: (a) sphere with demagnetization factors $N_x = N_y = N_z = 1/3$, (b)  rod with $N_z = 0$ and $N_x = N_y = 1/2$, (c) rod with $N_x = 0$ and $N_y = N_z = 1/2$, (d) thin platelet with $N_z = 1$ and $N_x =N_y = 0$. The dashed line in each panel indicates the resonance frequency in the paramagnetic limit $\omega=g\mu_{\rm B}\mu_0H/\hbar$.
}
\label{Fig:HelixResonances}
\end{figure}
There are two eigenmodes denoted by $\pm Q$ in the literature \cite{2012OnosePhysRevLett,Schwarze}.
Their frequencies are given by
\begin{eqnarray}
\hspace*{-2cm} 
\frac{\hbar \omega_{\pm Q}}{g \mu_B \mu_0 H_{c2}} =
\frac{1}{2(1+N_z \chi^{\rm int}_{\rm con})} \Big(
(2+\chi^{\rm int}_{\rm con}) (4 + (N_x + N_y) \chi^{\rm int}_{\rm con})
\\\nonumber
\hspace*{-2cm} + h^2 (-4 + \chi^{\rm int}_{\rm con} (-4 - N_y \chi^{\rm int}_{\rm con} + N_x (-1 + 2 N_y) \chi^{\rm int}_{\rm con} )) \pm
\\\nonumber
\hspace*{-2cm} \chi^{\rm int}_{\rm con} \Big[N_y^2 (2+\chi^{\rm int}_{\rm con} - h^2 \chi^{\rm int}_{\rm con})^2 + N_x^2 (2 + (1+h^2 (-1 + 2 N_y))\chi^{\rm int}_{\rm con} )^2
- 2 N_x N_y ((2+\chi^{\rm int}_{\rm con})^2
\\\nonumber
\hspace*{-2cm} - 2 h^2(2+\chi^{\rm int}_{\rm con}) (4+\chi^{\rm int}_{\rm con} +N_y \chi^{\rm int}_{\rm con})
+h^4 (8+\chi^{\rm int}_{\rm con}(8+\chi^{\rm int}_{\rm con}+2 N_y \chi^{\rm int}_{\rm con})))\Big]^{1/2} \Big)^{1/2}
\end{eqnarray}
where we expressed the energy $\mathcal{D} Q^2 = g \mu_B \mu_0 H^{\rm int}_{c2}$ in terms of the second critical field $H_{c2} = H^{\rm int}_{c2}(1+N_z \chi^{\rm int}_{\rm con})$, and we abbreviated $h = \cos\theta = H/H_{c2} = H_{\rm int}/H^{\rm int}_{c2}$. This result was first presented in Ref.~\cite{Schwarze} where the formula is given in terms of the susceptibility $\chi^{-1}_{\rm con} = (\chi^{\rm int}_{\rm con})^{-1} + N_z$.

For generic sample shapes eigenfrequencies of modes +Q, -Q
for $H<H_{c2}$
as well as the uniform Kittel mode of Eq.~(\ref{Kittel}) for $H>H_{c2}$ in the FP phase are shown in Fig.~\ref{Fig:HelixResonances}. For $N_x = N_y \neq 0$ the modes +Q and -Q are found to be degenerate at zero field $H=0$ and to split when $H$ approaches $H_{c2}$ [Fig.~\ref{Fig:HelixResonances} (a) and (b)]. In Fig.~\ref{Fig:HelixResonances} (c) the modes +Q and -Q exhibit different frequencies for all $H$ when $N_x \neq N_y$. In the limit of a thin platelet, $N_x = N_y = 0$ and $N_z = 1$, the eigenfrequencies are degenerate
for all fields $H < H_{c2}$, see Fig.~\ref{Fig:HelixResonances}(d). In this limit they are given by Kataoka's formula \cite{Kataoka1987}
\begin{eqnarray} \label{Kataoka}
\hbar \omega_{\pm Q}|_{\rm disc} = g \mu_B \mu_0 H^{\rm int}_{c2} \sqrt{1 + (1+\chi^{\rm int}_{\rm con})(1-h^2)}
\end{eqnarray}
with $h = H_{\rm int}/H^{\rm int}_{c2}$. This formula is consistent with the bulk spectrum of Eq.~(\ref{SpectrumHelix}) at $k_z = Q$. The branch in the FP phase is found either above or below the paramagnetic limit (dashed lines) depending on whether the shape anisotropy field adds to or subtracts from the applied field, respectively \cite{Gurevich}.

\subsection*{Polarization and spectral weights of modes}\label{PolWeights}

\begin{figure}
\centering
\includegraphics[width=0.9\textwidth]{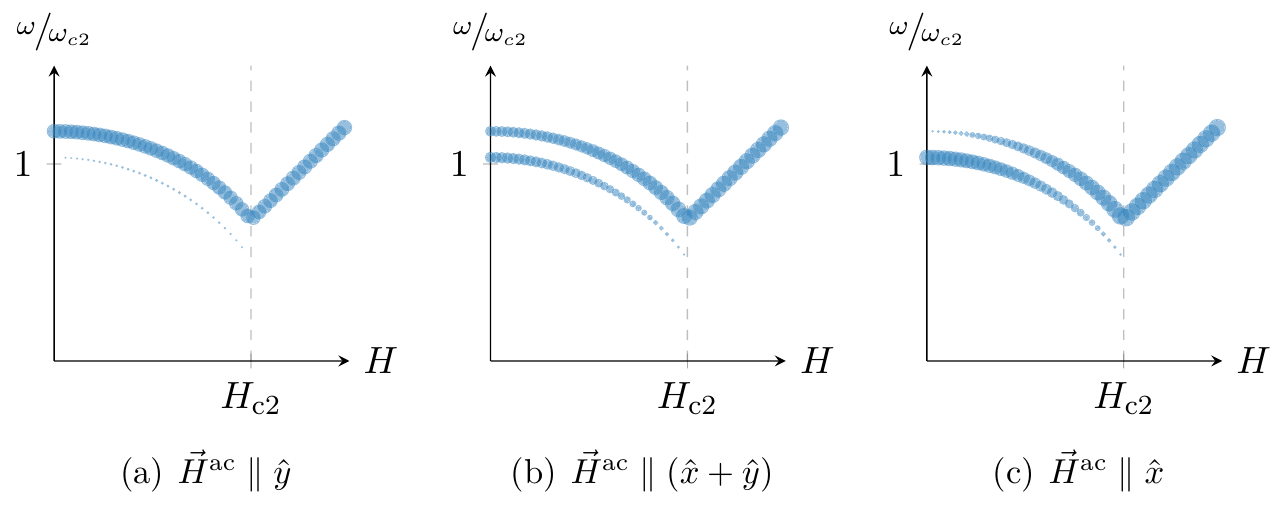}
\caption{Theoretically evaluated weights (symbol size) of the uniform resonances of the chiral magnet Cu$_2$OSeO$_3$ with $\chi_{\rm con}^{\rm int} = 1.76$ in the conical and field-polarized phase for a linearly polarized ac magnetic field. The shape is chosen to be rod-like with demagnetization factors $N_x = 0.07$, $N_y = 0.4$, and $N_z = 0.53$ and the dc field is applied along the $z$-axis. The weights depend on the relative orientation of the rod and the polarization of $\vec H_{ac}$. Generally, the weight of the $-Q$ mode vanishes at $H_{c2}$ while the weight of the $+Q$ mode smoothly connects to the Kittel mode.
Note that at $H=0$ the two modes $\pm Q$ can be selectively addressed by varying the polarization.
}
\label{Fig:HelixResonancesWeights}
\end{figure}

The eigenvectors of Eq.~(\ref{UniformModeHelix}) determine the helicity, the polarization and weights of the resonances. The weights as a function of field are shown in Fig.~\ref{Fig:HelixResonancesWeights}. The $\pm Q$ modes have a well-defined helicity provided that they are non-degenerate. The uniform magnetization oscillates counterclockwise for the $+Q$ mode and clockwise for the $-Q$ mode. As a result, the weight of the $+Q$ mode continuously connects to the Kittel mode at the critical field, $H_{c2}$, because they possess the same helicity, i.e., for both modes the uniform magnetization oscillates counterclockwise. In contrast, the weight of the $-Q$ mode vanishes as $H_{c2}$ is approached.

\begin{figure}
\centering
\includegraphics[width=0.5\textwidth]{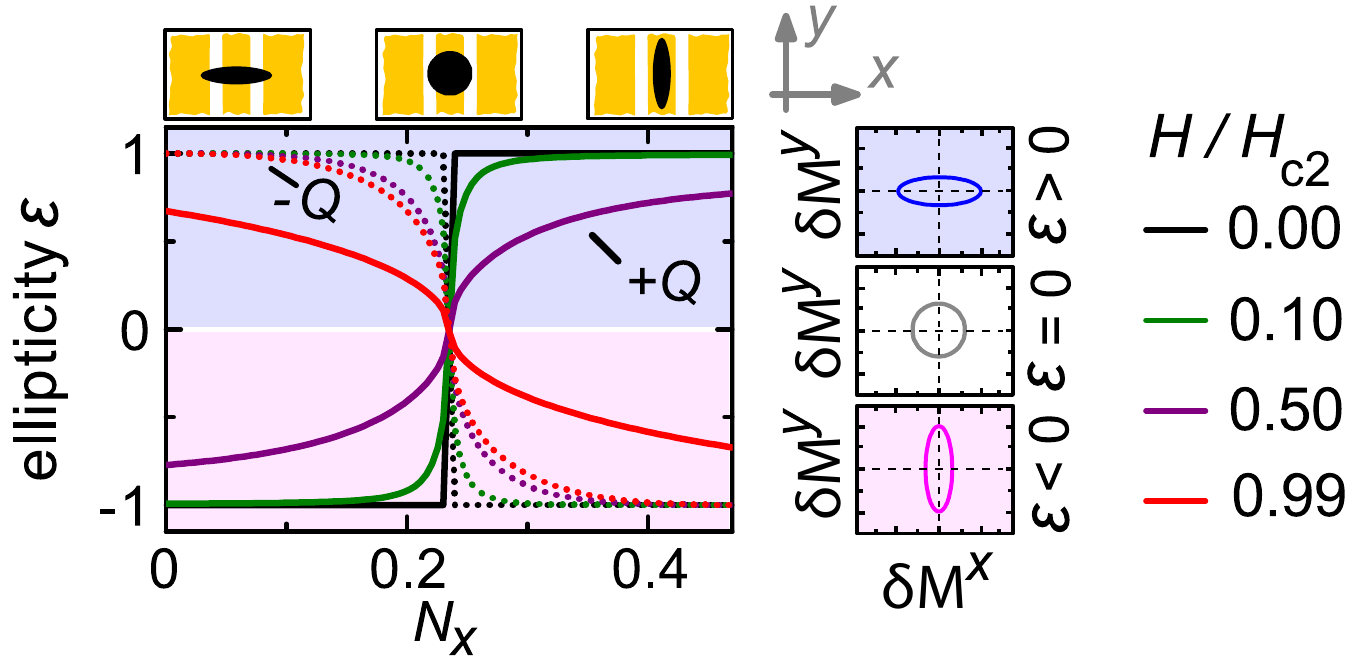}
\caption{Ellipticity as defined in Eq.~(\ref{Ellipticity}) of the uniform resonances $\pm Q$ in the conical phase as a function of demagnetization factor $N_x$ at fixed $N_z = 0.53$ and $N_y = 1 - N_x - N_z$ for various magnetic fields and $\chi_{\rm con}^{\rm int} = 1.76$ \cite{Stasinopoulos2016}. At zero field $H=0$, the $\pm Q$ modes are linearly polarized $|\varepsilon| = 1$ for generic sample shapes.
 }
\label{Fig:Ellipticity}
\end{figure}

If the modes are degenerate as for $N_x = N_y = 0$, one can always choose a basis of, for example, two circularly polarized degenerate eigenmodes. However, in the generic case of two non-degenerate modes, each of them possesses a distinct polarization and ellipticity. This leads to an interesting dependence of their weights on the orientation $\vec H_{ac}$ of a linearly polarized ac field as illustrated in Fig.~\ref{Fig:HelixResonancesWeights}.
Both modes are in general elliptically polarized where the axes are defined by the sample shape. We introduce the ellipticity defined by
\begin{equation} \label{Ellipticity}
\varepsilon = {\rm sign}\{\delta M_x - \delta M_y\} \frac{\sqrt{|\delta M^2_x - \delta M^2_y|}}{{\rm max}\{\delta M_x, \delta M_y\}}
\end{equation}
where $\delta M_{x,y}$ are the positive amplitudes of the oscillations within the $(x,y)$ plane. We introduce a sign so that $\varepsilon > 0$ if the major axis is aligned with the $x$-axis and $\varepsilon < 0$ if it is aligned with the $y$-axis. The ellipticity for the $\pm Q$ modes is shown in Fig.~\ref{Fig:Ellipticity} for a specific value of $N_z$ as a function of $N_x$. Interestingly, we find that at zero field $\vec H = 0$ the two modes are generically linearly polarized, i.e., $\varepsilon = \pm 1$.

The linear polarization is related to the $\pi$-rotation symmetry of the helix that is present at zero field, see the discussion below Eq.~(\ref{ConHelix}). This symmetry is represented by the matrix operator
\begin{equation} \label{PiRot}
\left(\begin{array}{cc}
0 & \mathds{1} \\ \mathds{1} & 0
\end{array}\right)
\end{equation}
that commutes at $\vec H = 0$ with the Hamiltonian, i.e., with the
$4\times 4$ matrix of the wave equation (\ref{UniformModeHelix}). As a result, the two eigenmodes $\pm Q$ are also eigenvectors of the matrix (\ref{PiRot}) with two distinct eigenvalues. Moreover, the amplitudes $\delta M_{x}$ and $\delta M_{y}$  are determined by the projection of the wavefunction $(\vec \Psi^T(\vec Q,\omega), \vec \Psi^T(-\vec Q,\omega))$, see Eq.~(\ref{HelixUniformModeXY}), onto different degenerate subspaces of the $\pi$-symmetry operator (\ref{PiRot}). At zero field, either of the two amplitudes thus vanishes leading to a linear polarization. In less mathematical terms, the $\pi$-rotation symmetry ensures that there are always pairs of spins within the helix whose local precessions conspire such that in total only a linear polarization remains. This situation seems to be similar to easy-plane antiferromagnets where the precession of magnetic moments on the two sublattices also combine to yield a linearly polarized uniform mode at antiferromagnetic resonance \cite{Sievers::PhysRev::1963,Gurevich}. In contrast to antiferromagnets, the linear polarization for chiral magnets is however an interplay between DMI and sample shape. Because the relevant energy scale of DMI is orders of magnitude smaller compared to antiferromagnetic exchange interaction, the linearly polarized modes of \cso\ reside at small frequencies \cite{Gurevich}.

Figure~\ref{Fig:Ellipticity} also indicates that the ellipticity of the $+Q$ mode changes sign implying that it changes the major axis of its elliptical polarization as a function of field. Generally, this is expected to occur at the magnetic field
\begin{eqnarray}
\hspace{-2.5cm} \frac{H^{\rm circ}_+}{H_{c2}} =
\frac{2+\chi_{\rm con}^{\rm int}}
{\textstyle
\sqrt{(2+\chi_{\rm con}^{\rm int})(2+(2-N_z)\chi_{\rm con}^{\rm int}) +
\sqrt{(1-N_z)\chi_{\rm con}^{\rm int}(2+\chi_{\rm con}^{\rm int})^2(4+(1-N_z)\chi_{\rm con}^{\rm int})}}}.
\end{eqnarray}
For Cu$_2$OSeO$_3$ with $\chi_{\rm con}^{\rm int} = 1.76$ and a sample with $N_z = 0.53$ this yields $H^{\rm circ}_+/H_{c2} \approx 0.76$. At this specific field, the $+Q$ mode is circularly polarized irrespective of the values of the demagnetization factors $N_x$ and $N_y$.

In the early experiments by Date {\it et al.} \cite{Date1977} both $\pm Q$ excitation modes were detected on MnSi. Onose {\it et al.} \cite{2012OnosePhysRevLett} reported the two modes for Cu$_2$OSeO$_3$ (Fig. \ref{figOnose}). Finally, Schwarze {\it et al.} \cite{Schwarze} performed microwave resonance experiments on MnSi, Cu$_2$OSeO$_3$ and Fe$_{0.8}$Co$_{0.2}$Si, and found quantitative agreement with the theory reviewed above. The linear polarization of the $\pm Q$ modes in zero field was explored and experimentally confirmed by Stasinopoulos {\it et al.} \cite{Stasinopoulos2016}.

\subsection{Magnon excitation of the skyrmion crystal}

The Dzyaloshinskii-Moriya interaction not only favours helices but also topological solitons.
Within the field-polarized state, a single soliton is generated by the map
\begin{equation} \label{Skyrmion}
\hat n_{\rm Sk}(\vec r) = e^{i \vec f_{\rm Sk}(\vec r) \vec {\bf L}} \hat z =
\left(\begin{array}{c}
- \sin \chi \sin \theta(\rho) \\
\cos \chi \sin \theta(\rho)\\
\cos \theta(\rho)
\end{array} \right)
\end{equation}
where $\vec {\bf L}$ is the spin-1 operator of Eq.~(\ref{Spin1}) and $\vec f_{\rm Sk}(\vec r) = \theta(\rho) \hat \rho$ with $\vec r = (\rho \cos \chi, \rho \sin \chi, z)$ in cylindrical coordinates and $\hat \rho = (\cos \chi, \sin \chi, 0)$. Note that $\vec f_{\rm Sk}$ is a conservative vector field and possesses a potential according to $\vec f_{\rm Sk}(\vec r) = \vec \nabla \Theta_{\rm Sk}(\rho)$ with $\theta(\rho) = \partial_\rho \Theta_{\rm Sk}(\rho)$. The function $\theta(\rho)$ obeys the asymptotics $\theta(0) = \pi$ and $\theta(\rho) \to 0$ as $\rho \to \infty$. This soliton is translational invariant along the field-direction, i.e., the $z$-axis, and it varies within the plane perpendicular to the field. In a three dimensional bulk magnet, it defines in fact a string.
The soliton carries a finite topological skyrmion number so that it is referred to simply as a skyrmion [Fig. \ref{fig2} (d)]. The topological skyrmion density within the $(x,y)$ plane is defined by $\rho_{\rm top} = \frac{1}{4\pi} \hat n (\partial_x \hat n \times \partial_y \hat n)$. Integrating the topological density of the texture (\ref{Skyrmion}) yields an integer, $\int dx dy \rho_{\rm top} = -1$.
Minimizing the exchange energy of Eq.~(\ref{F0}) with the Ansatz (\ref{Skyrmion}), i.e., neglecting the dipolar interactions one obtains an ordinary differential equation that determines the function $\theta(\theta)$, which was first discussed by Bogdanov and Hubert \cite{Boganov1994}. Importantly, one finds that $\theta(\rho)$ decays exponentially so that
the skyrmion is confined to an area that scales approximately as $1/\vec H^2$ where $\vec H$ is the magnetic field \cite{Leonov2016}.

The magnon scattering off such a single skyrmion will be discussed in section \ref{sec:MagSkyScattering}. Here, we focus on the spin wave excitation of a skyrmion crystal. When it is energetically advantageous to condense skyrmions, they proliferate and form a skyrmion lattice  [Fig.~\ref{fig2}(e)]. Such a magnetic state is observed in bulk crystals of chiral magnets at intermediate fields and close to the critical temperature [Fig.~\ref{fig2}(a)]. A skyrmion crystal state can be generated with the help of the map
\begin{equation} \label{SkX}
\hat n_{\rm SkX}(\vec r) = e^{i \vec f_{\rm SkX}(\vec r) \vec {\bf L}} \hat z
\end{equation}
with $\vec {\bf L}$ of Eq.~(\ref{Spin1}). The potential of the conservative vector function $\vec f_{\rm SkX}(\vec r) = \vec \nabla \Theta_{\rm SkX}(\vec r_\perp)$ again only varies within the plane perpendicular to the magnetic field, $\vec r_\perp = (x,y,0)$. It possesses the periodicity of a two-dimensional hexagonal Bravais lattice $L$, i.e., $\Theta_{\rm SkX}(\vec r_\perp) = \Theta_{\rm SkX}(\vec r_\perp + \vec R_i)$ for any vector $\vec R_i \in L$. If the lattice constant $a$ is large compared to the skyrmion radius, this potential can be approximated to be the sum of individual skyrmion potentials $\Theta_{\rm SkX}(\vec r_\perp) \approx \sum_{\vec R_i \in L} \Theta_{\rm Sk}(\vec r_\perp - \vec R_i)$. However, it turns out that the skyrmions are closely packed, and the lattice constant $a$ is in fact comparable to the skyrmion diameter so that $\Theta_{\rm Sk}$ is expected to differ from the single-skyrmion solution.
In any case, minimization of the free energy (\ref{F0}) with such a type of Ansatz does not yield a smaller energy than the competing conical state of Eq.~(\ref{ConHelix}). The phase diagram of Fig.~\ref{fig2} can therefore not be explained on the mean-field level. It was argued in Ref.~\cite{Muehlbauer2009} that fluctuations around the mean-field potential stabilizes the SkL, which was subsequently confirmed by Monte-Carlo simulations \cite{Buhrandt:2013}.

Once the SkL is stabilized the magnon spectrum can be derived on the mean-field level taking into account both contributions, the exchange and the dipolar energies of Eqs.~(\ref{F0}) and (\ref{Fdip}), respectively. However, for practical purposes it is more convenient to relax the constraint of $\hat n$ being a unit vector and to work instead within the framework of a linear sigma model. In the following, we present numerical results for the spin wave spectrum of skyrmion crystals in bulk chiral magnets. For computational details we refer the reader to Refs.~\cite{Schwarze} and \cite{Waizner2017}.

\subsubsection{Spin waves of the skyrmion crystal in bulk samples: finite wavevectors}

\begin{figure}
\centering
\includegraphics[width=\textwidth]{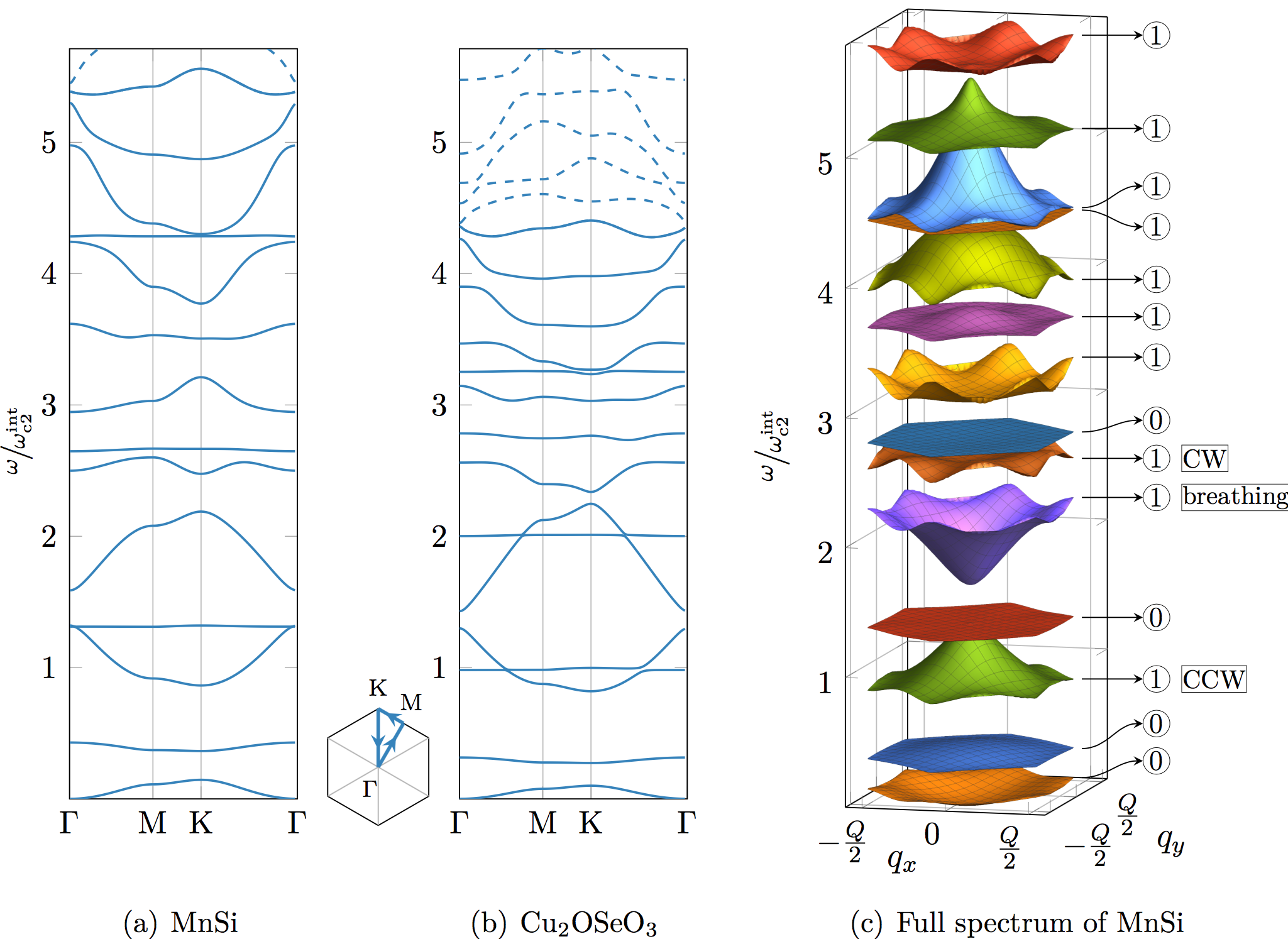}
\caption{Magnon band structure of the hexagonal skyrmion crystal in chiral magnets within the first Brillouin zone in unit of $\hbar \omega^{\rm int}_{c2} = g \mu_B \mu_0 H_{c2}^{\rm int}$. The magnon wavenumber is confined to the two-dimensional plane orthogonal to the magnetic field, i.e. $k_z = 0$. Panel (a) and (c) show the spectrum for the parameter $\chi_{\rm con}^{\rm int} = 0.34$ corresponding to MnSi and panel (b) for Cu$_2$OSeO$_3$ with $\chi_{\rm con}^{\rm int} = 1.76$. The inset between (a) and (b) displays the first Brillouin zone. The Chern number of the bands are given on the right-hand side of panel (c).  The magnon excitations at the zone center of the 3rd, 5th and 6th band in panel (c) correspond to the uniform CCW, breathing and CW  modes, see Figs.~\ref{Fig:SkXResonancesFieldDependence} and \ref{Fig:SkXResonancesIllustration}.
 }
\label{Fig:SkXBandStructure}
\end{figure}

The periodicity of the two-dimensional hexagonal skyrmion crystal gives rise to a magnon band structure with a two-dimensional Brillouin zone in the $(x,y)$ plane perpendicular to the applied magnetic field [inset in Fig.~\ref{Fig:SkXBandStructure} (a)]. Figure~\ref{Fig:SkXBandStructure} (a) shows the magnon dispersion for vanishing out-of-plane momentum $k_z = 0$ along particular symmetry directions within the first Brillouin zone for the parameter $\chi_{\rm con}^{\rm int} = 0.34$ corresponding to MnSi. Figure~\ref{Fig:SkXBandStructure} (b) shows the spectrum for Cu$_2$OSeO$_3$ with $\chi_{\rm con}^{\rm int} = 1.76$ yielding a compressed band structure as compared to panel (a). For better comparison, the first 14 bands in (a) and (b) are shown as solid lines and the remaining bands are displayed as dashed lines. Figure~\ref{Fig:SkXBandStructure} (c) shows the complete magnon spectrum of MnSi for wavevectors in the plane of the SkL.

Similarly to the conical helix, the skyrmion crystal breaks translational invariance so that the spectrum possesses a Goldstone mode. This mode arises at the $\Gamma$ point where the lowest band touches zero energy. Its excitation energy vanishes quadratically with momentum, $\omega \sim k^2$, which has been attributed to the topological nature of the skyrmions in Refs.~\cite{Petrova2011,Zang2011}.

The non-trivial topology has further consequences.
As will be explained in detail in section \ref{sec:MagSkyScattering}, each skyrmion acts like
an orbital magnetic field with quantized flux resulting in skew scattering.
It is therefore expected that a magnon in a skyrmion crystal experiences this magnetic flux per magnetic unit cell. As a consequence, magnons should occupy Landau levels that are reflected in non-trivial Chern numbers of the band structure \cite{Hasan2010}. The Chern numbers of the magnon bands for $\chi_{\rm con}^{\rm int} = 0.34$ are shown on the right-hand side of Fig.~\ref{Fig:SkXBandStructure}(c). We find that for the lowest 14 bands, for which we were able to compute these numbers reliably, most of the bands have a Chern number 1 with a few exceptions, in particular, at low energies where the Chern number is zero.
For the parameters of MnSi, the magnon bands are sufficiently well separated allowing for an unambiguous computation. For the parameters of Cu$_2$OSeO$_3$ some of the bands come very close rendering the computation of Chern numbers difficult. Up to such ambiguities, the Chern numbers listed in Fig. \ref{Fig:SkXBandStructure} (c) are consistent with the work of Rold\'an-Molina {\it et al.} \cite{Roldan-Molina2016}. The band structures of Fig. \ref{Fig:SkXBandStructure} will be further discussed in sect. \ref{prospects}.

\subsubsection{Spin waves of the skyrmion crystal in bulk samples: uniform mode}

\begin{figure}
\centering
\includegraphics[width=0.5\textwidth]{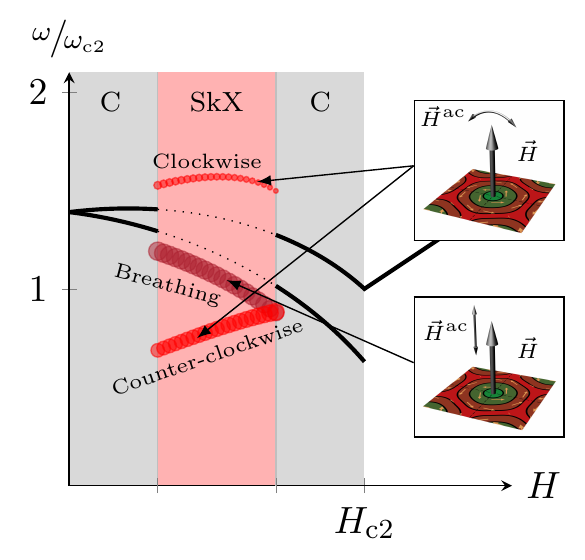}
\caption{Field dependence of the resonance frequencies of the breathing mode, counterclockwise (CCW) and clockwise (CW) mode -- the three uniform resonances of the skyrmion crystal -- as expected for a spherical sample of Cu$_2$OSeO$_3$.
An ac magnetic field linearly polarized within the skyrmion crystal plane excites both CW and CCW modes where the latter always possesses larger weight as indicated by the size of the dots. The breathing mode possesses an out-of-plane linear polarization. The black solid lines indicate the positions of the resonances of the other phases.
}
\label{Fig:SkXResonancesFieldDependence}
\end{figure}

Due to the back-folding induced by the periodicity of the Skyrmion crystal, there is a multitude of modes present at the $\Gamma$ point, i.e., at zero wavevector. However, similar to the conical helix, only a few of them are magnetically active and excited by an homogeneously oscillating magnetic field. For the skyrmion crystal, there are three uniform magnetic modes, breathing, CCW and CW mode, first identified theoretically by Mochizuki \cite{2012MochizukiPhysRevLett}.

\begin{figure}
\centering
\includegraphics[width=0.45\textwidth]{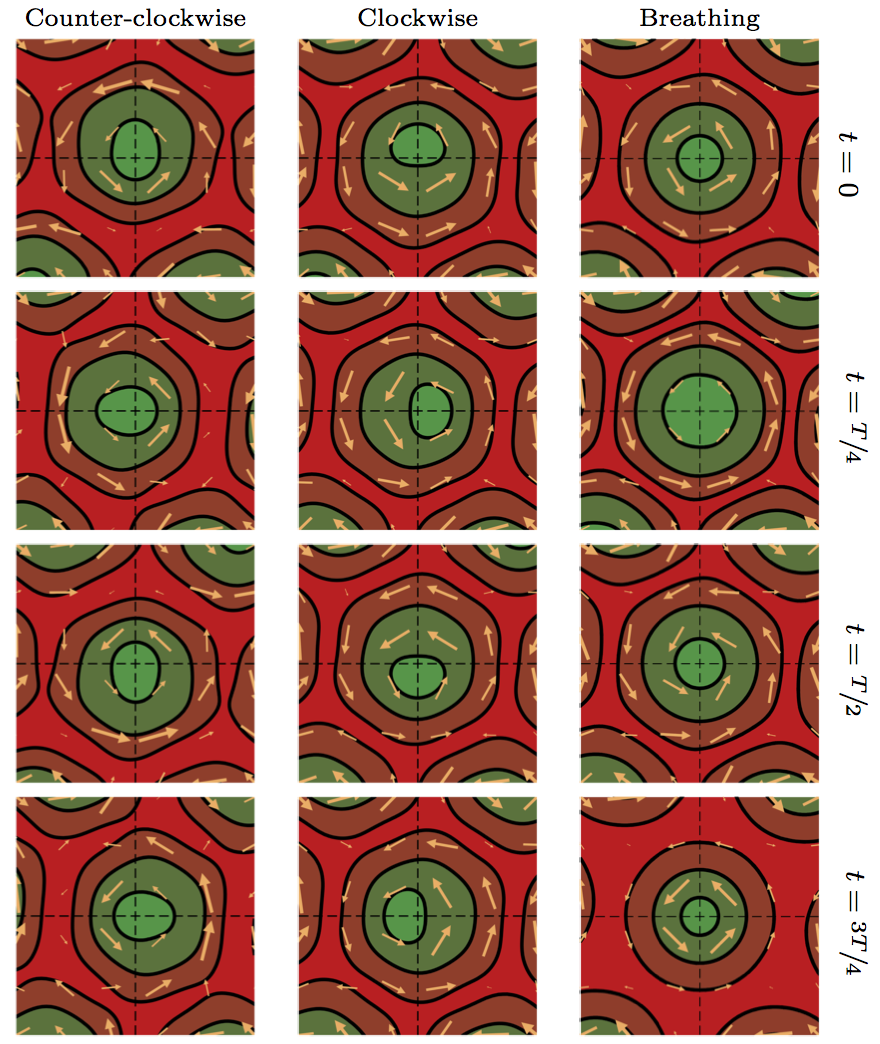}
\caption{Evolution of the magnon wavefunction in real space as a function of time for the three uniform resonance modes of the skyrmion crystal. T is the respective time period.
The dc magnetic field points out of the image plane. The in-plane components of the magnetization are shown by amber colored arrows while the out-of-plane component is represented by a contour plot where green corresponds to the magnetization pointing anti-parallel to the dc field, i.e., into the plane, and red indicates magnetization pointing out of plane.
}
\label{Fig:SkXResonancesIllustration}
\end{figure}

The breathing mode possesses a macroscopically dipole moment oscillating out-of-plane, i.e., it is linearly polarized and excited with a longitudinal ac magnetic field (Fig.~\ref{Fig:SkXResonancesFieldDependence}). The size of each skyrmion in the crystal performs an oscillating motion as illustrated in Fig.~\ref{Fig:SkXResonancesIllustration}.
Its resonance frequency decreases with magnetic field similar to the $\pm Q$ modes of the adjacent conical phase. In case of phase coexistence within the sample, the breathing mode can be distinguished by its polarization which is orthogonal to the ones of the $\pm Q$ modes.

The uniform magnetization of the CCW and CW modes oscillates counterclockwise and clockwise, respectively, in the plane of the skyrmion crystal. They are excited with an in-plane ac field.
These modes are in general elliptically polarized along different axes that are determined by the sample shape. Their ellipticity however varies only slightly within the limited magnetic field range where the skyrmion crystal is found \cite{Waizner2017}. Generally, an ac field linearly polarized within the plane excites both modes but the CCW mode always has a larger weight (Fig.~\ref{Fig:SkXResonancesFieldDependence}).
Whereas the resonance frequency of the CCW increases with field, the CW frequency only depends weakly on the magnetic field consistent with experimental observations \cite{Schwarze,2012OnosePhysRevLett}.

\section{Prospects and challenges}\label{prospects}

After reviewing the spin wave excitations of the various magnetic phases in bulk samples of cubic chiral magnets, we discuss future research avenues as well as challenges for magnonics applications in the following.

\subsection{Magnonics with cubic chiral magnets}

The chiral magnets with bulk DMI
provide strikingly new characteristics in view of magnonic crystals and manipulation of spin waves in solids. The DMI-induced spin helix introduces a periodic potential [Eq. (\ref{H0Helix})] for magnons that leads to backfolding of spin-wave dispersion relations and Brillouin zone boundaries at half the pitch vector $Q$ (Fig. \ref{Fig:HelixSpectrum}). For \cso\ the intrinsic pitch length that defines the periodicity of the
helix amounts to about 60 nm. Thereby magnonic crystal equivalent band structures are formed avoiding any nanopatterning that typically introduces roughness and inhomogeneous broadening. In addition, in the helical phase, the band structure changes qualitatively as a function of momentum $\vec k_\perp$ transverse to the helix axis, see Fig.~\ref{Fig:HelixSpectrum}. For large $|\vec k_\perp| \gg Q$, the lowest bands become flat implying that the magnons are localized along the helix axis and only propagate transverse to it, i.e., the spin waves experience a channeling effect. Channeling of spin waves is well known for two-dimensional magnonic crystals made from antidot lattices \cite{Neusser2008,Neu2010}. In their case the inhomogeneous demagnetization field induced by in-plane magnetic fields creates narrow channels that are perpendicular with respect to the applied field direction. In the helical or conical phase, spin waves are localized within planes perpendicular to the pitch vector $\vec{Q}$. The orientation of the latter can be manipulated with an applied magnetic field allowing to control the channeling. Future research might explore the multitude of tuneable parallel channels in view of spin wave multiplexer on the nanoscale.

\begin{figure}
\centering
    \includegraphics[width=\textwidth]{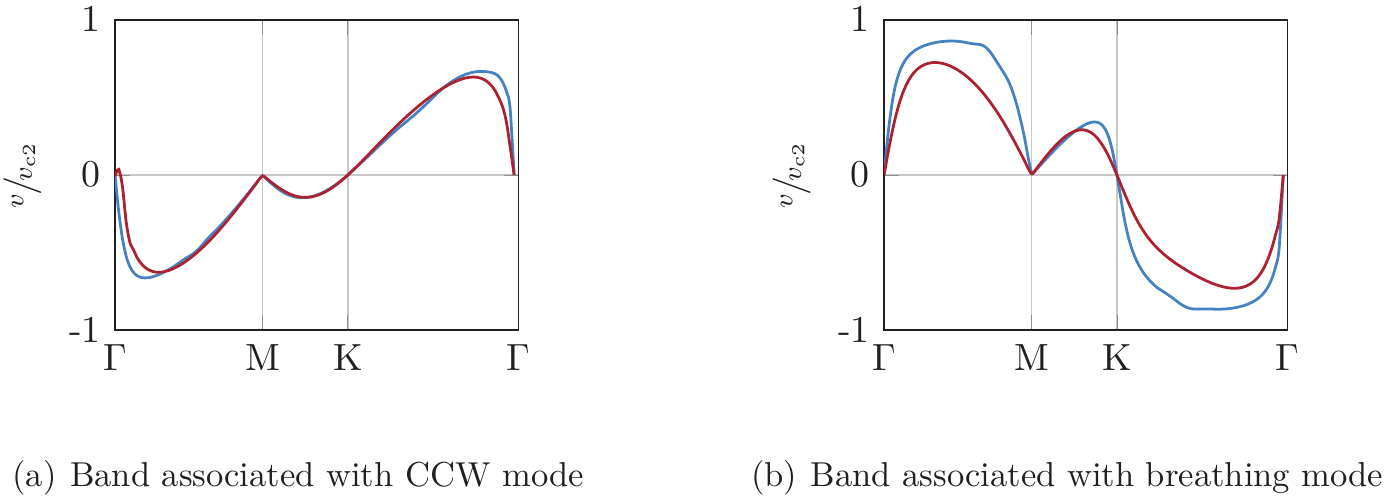}
	\caption{Magnon group velocity $v = \partial_{k} \hbar \omega_{\vec k}$ along certain directions within the first Brillouin zone of the skyrmion lattice  for MnSi (blue solid line) and Cu$_2$OSeO$_3$ (red solid line). The two bands are shown that contain at their respective $\Gamma$ point the magnetically active breathing and CCW modes, see Fig.~\ref{Fig:SkXBandStructure}. The velocities are on the order of $v_{c2} = \omega_{c2}/Q = g \mu_B \mu_0 H^{\rm int}_{c2}/(\hbar Q)$.
	}
	\label{fig:groupvelocity}
\end{figure}

Within the phase of the skyrmion lattice, there also exist dispersive magnon bands. The dispersion is particularly pronounced for the magnetically active bands associated with the breathing and CCW modes at their respective $\Gamma$ point, see Fig.~\ref{Fig:SkXBandStructure}. Their group velocities $v = \partial_{k} \hbar \omega_{\vec k}$ along certain lines within the two-dimensional Brillouin zone are shown in Fig.~\ref{fig:groupvelocity}. The velocities are on the order of $v_{c2} = \hbar \omega_{c2}/(\hbar Q)$ which corresponds to $v_{c2} \approx 300$ m/s for MnSi and $v_{c2} \approx 140$ m/s for Cu$_2$OSeO$_3$.

The Dzyaloshinskii-Moriya interaction imposes boundary condition on surfaces that result in a surface twist. In the FP phase, this leads to exchange spin waves that are localized to the surfaces of the sample
\cite{Garcia-Sanchez2014,Mueller2016}. They will be further discussed below in section \ref{sec:SpinWavesThinFilms}. In the SkL phase, surface spin-wave modes are also expected for a different reason. We have discussed in Fig.~\ref{Fig:SkXBandStructure} that some of the magnon bands possess a non-trivial Chern number resulting from the non-trivial topology of the magnetic texture. According to the bulk-boundary correspondence, this implies the presence of topologically protected magnon edge states. Such magnon edge states have been previously discussed in a different context for the pyrochlore Lu$_2$V$_2$O$_7$ \cite{LifaZhang2013}. Moreover, the magnon bands with a finite Chern number should contribute to a magnon Hall effect, i.e., spin accumulation transverse to the magnon flow \cite{Mochizuki2014,Iwasaki2014,Schuette2014}.

Periodically patterned ferromagnetic materials allow one to create magnonic crystals with reconfigurable wave properties. Via different magnetic states one-and-the-same magnonic crystal can exhibit different forbidden frequency gaps and Brillouin zone boundaries (for a review see Ref. \cite{KrawcykGrundler2014}). To control the characteristics different techniques have been explored or suggested such as e.g. magnetic fields \cite{JTopp2010,Tacchi2010,Haldar2016}, current pulses \cite{chumak2010}, light patterns \cite{Vogel2015}, thermally assisted scanning probe lithography \cite{Albisetti} and electric fields \cite{Brandl201413}. Technologically electric field control is most advantageous as electric fields can be confined very precisely and allow for a low power consumption. The SkL in \cso\ has already been shown to get distorted and rotate under the application of an electric field due to magnetoelectric coupling \cite{2012WhiteJPhysCondensMatter,White2014}. Electric-field controlled magnetic crystals based on SkLs should thus be possible. At the same time, the SkL might be exploited as either a resonant or non-resonant magnonic grating coupler if integrated to a coplanar waveguide \cite{Yu2016,Yu2013}. Relevant wavelengths for emitted spin waves are on the order of the inverse of reciprocal lattice vectors $G$ \cite{Yu2016,Yu2013}. In case of an SkL we assume $G=Q$ leading to spin-wave wavelengths $\lambda=2\pi/G=2\pi/Q$ on the order of 20 to 70 nm for the materials listed in Tab. \ref{Table1}.
Applying an electric field would rotate the orientation of the hexagonal magnetic grating around the field direction and allow to emit short-wavelength spin waves in different and tuneable directions transverse to $\vec{H}$. The continuous tunability of propagation directions has not yet been foreseen with conventional ferromagnets. By tilting the magnetic field $\vec{H}$ stabilizing the SkL the grating coupler can be further adjusted and becomes versatile for even more spatial directions.

In metallic materials applied currents have been shown to couple efficiently to an SkL. At very small current densities an SkL is depinned and shifted in space via spin-transfer torque (STT) \cite{2010JonietzScience,2012YuNatCommun}. This effect allows for the current-controlled positioning of skyrmion-based magnonic crystals. In a moving SkL Doppler-shift like frequency variations of spin waves might be anticipated \cite{Vlaminck410}.

\subsection{Shape- and orientation dependent effects}

The resonance frequencies of the uniform $\pm Q$ modes of the conical helix strongly depend on the shape of the sample as illustrated in Fig.~\ref{Fig:HelixResonances}. For the breathing, CW and CCW modes of the skyrmion lattice the variation of eigenfrequencies with demagnetization factors is less pronounced \cite{Schwarze}. Still the order of magnon bands in the skyrmion lattice varies with the strength of the effective dipolar interaction as shown for MnSi and Cu$_2$OSeO$_3$ in Fig.~\ref{Fig:SkXBandStructure}.

The three uniform excitation modes of the skyrmion lattice have been also observed for the N{\'e}el-type skyrmion lattice in the rhombohedral GaV$_4$S$_8$ \cite{Ehlers2016a,Ehlers2016b}, which is characterized by a uniaxial anisotropy. However, here the breathing mode was experimentally found to possess a smaller resonance frequency than the CCW mode, i.e., the hierarchy of resonances differs from the one of the cubic chiral magnets in Fig.~\ref{Fig:SkXResonancesIllustration}. This has also been theoretically confirmed by numerical simulations in Ref.~\cite{Zhang2016}. Moreover, the eigenfrequencies and set of modes in this material class are expected to vary distinctly if an applied magnetic field is misaligned with respect to the easy axis. The misaligned field leads to a distortion of the skyrmion lattice with the concomitant change in resonance frequencies.

The interplay of bulk DMI and shape anisotropy leads to a linear polarization of the magnetization dynamics in the spin-helix state at small magnetic field,
see section \ref{PolWeights}. Relevant eigenfrequencies are in the few GHz frequency regime for which transmission lines and coplanar waveguides are impedance matched and guide linearly-polarized microwaves efficiently on microchips. In conventional ferro- and ferrimagnets such linearly polarized spin-precessional motion is not accomplished. For them the spin-precessional motion is elliptically or circularly polarized, and its polarization does not match with the conventional microwave technology. Helical modes of chiral magnets with bulk DMI provide a better matching concerning polarization planes.

\subsection{Magnetochiral and multiferroic characteristics}
The magnetic skyrmions in \cso\ are accompanied by electric dipoles \cite{0953-8984-27-50-503001}. Hence, magnons are present that exhibit both magnetic and electric activities which undergo interference effects and lead to directional dichroism. The relevant electric susceptibilities of the rotational and breathing modes in the SkL were reviewed in Ref. \cite{0953-8984-27-50-503001}. Linearly polarized microwaves guided through an SkL experience directional dichroism in both configurations for which the wavevector $\vec{k}_{\rm em}$ of the electromagnetic wave is either perpendicular or parallel to the magnetic field $\vec{H}$ stabilizing the SkL. Counterpropagating electromagnetic waves encounter different absorption leading to nonreciprocal microwave properties (directional dichroism). For the case that $\vec{k}_{\rm em}$ is collinear with $\vec{H}$ the directional dichroism is called magnetochiral effect \cite{0953-8984-27-50-503001}. This effect has been found not only in the SkL of bulk \cso\ but also in the conical and field-polarized phases. This kind of microwave nonreciprocity can thus be exploited over a broad frequency regime. For a detailed discussion we refer the reader to Refs. \cite{2015Okamura,2015MochizukiPhysRevLett.114.197203}. To harvest such characteristics in microwave technology bulk materials with DMI are needed that are insulators and form Skyrmions at room temperature. While bulk metals with DMI and ultrathin metallic layers incorporating interfacial DMI provide magnetic skyrmions at room temperature \cite{Tokunaga2015,Soum2016}, to our knowledge, bulk insulators with room-temperature skyrmions have not yet been reported. Corresponding materials still need to be identified.

Interestingly micromagnetic simulations have suggested that a linearly polarized microwave field applied to a chiral magnet can displace both an individual skyrmion and an SkL in the presence of a symmetry-breaking magnetic field component. The maximum velocity is found for a frequency of the microwave that coincides with the resonance frequency of the breathing mode of the skyrmions \cite{PhysRevB.92.020403}.

\subsection{Magnonics with individual skyrmions}
\label{sec:MagSkyScattering}

\begin{figure}
\centering
\includegraphics[width=0.4\textwidth]{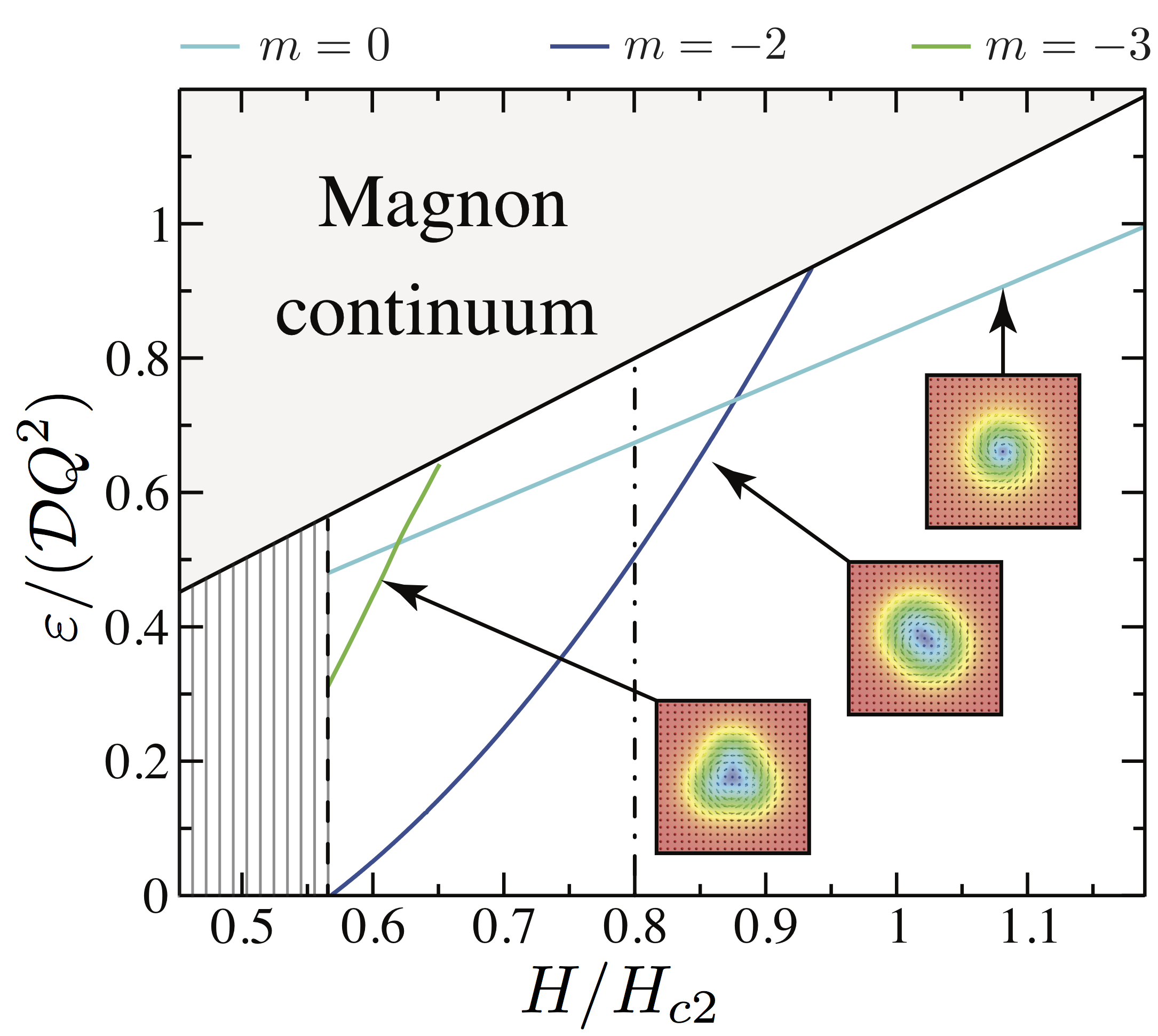}
\caption{Magnon spectrum of the field-polarized state in the presence of a single skyrmion \cite{Schuette2014}. There exist few magnon-skyrmion bound states below the magnon gap. The breathing mode parametrized by the angular momentum $m=0$ extends up to high fields. A quadrupolar mode $m=-2$ appears for smaller fields whose resonance frequency vanishes for $H/H_{c2} \approx 0.56$ indicating the elliptical instability of the skyrmion \cite{Boganov1994}. A sextupolar mode $m=-3$ is only realized in a small magnetic field range above the elliptical instability.
}
\label{Fig:SkyrmionSpectrum}
\end{figure}
For skyrmion-based spin-electronics an individual skyrmion formed in a ferromagnetic matrix is expected to be technologically relevant \cite{2013SampaioNatNanotech}.
In insulators, spin waves can then be used move the skyrmion, e.g., in a nanotrack \cite{Kong2013,Lin2014a,Kovalev2014,Mochizuki2014,Iwasaki2014,Schuette2014,XZhang2017}. For the functionalization of this magnon-skyrmion momentum-transfer, the understanding of the spin wave scattering off a skyrmion is essential.

When spin waves approach the skyrmion they scatter off it, and they can even form bound states.
This magnon-skyrmion scattering problem was considered in Refs.~\cite{Lin2014,Iwasaki2014,Schuette2014} on the level of the exchange energy of Eq.~(\ref{F0}). Here, we review the results of Ref.~\cite{Schuette2014} where the magnon Hamiltonian was derived analytically and solved numerically for a finite in-plane magnon momentum $\vec k_\perp = (k_x,k_y,0)$. The resulting spectrum (without uniaxial anisotropy) is shown in Fig.~\ref{Fig:SkyrmionSpectrum}. Apart from scattering states above the magnon gap, there exist a few in-gap states corresponding to magnon-skyrmion bound modes. Above the critical field $H_{c2}$, there is only a single breathing mode for which the skyrmion radius oscillates in time. An additional quadrupolar and sextupolar mode appears only for smaller fields where the field-polarized state in a three-dimensional bulk material is already unstable.

In a magnetic layer, the conical state is suppressed and the field-polarized state remains stable below $H_{c2}$. The vertical dashed-dotted line in Fig.~\ref{Fig:SkyrmionSpectrum} then indicates a global instability where the energy of skyrmions becomes negative triggering the formation of a skyrmion crystal. Below this field, the skyrmion in the metastable field-polarized phase becomes elliptically unstable \cite{Boganov1994} when the resonance frequency of the quadrupolar mode vanishes.

Only the breathing mode gives rise to an oscillating average magnetic dipole moment that can be excited with an ac magnetic field longitudinal to the dc field, i.e., along the $z$-axis.  In the insulator Cu$_2$OSeO$_3$ the quadrupolar mode is accompanied with an oscillating electric dipole moment and could therefore be electrically excited (electromagnon) \cite{Schuette2014}. We mention that the breathing mode is associated with an oscillating toroidal moment so that it can be excited also with a combination of in-plane magnetic and electric fields.

\begin{figure}
\centering
\includegraphics[width=0.7\textwidth]{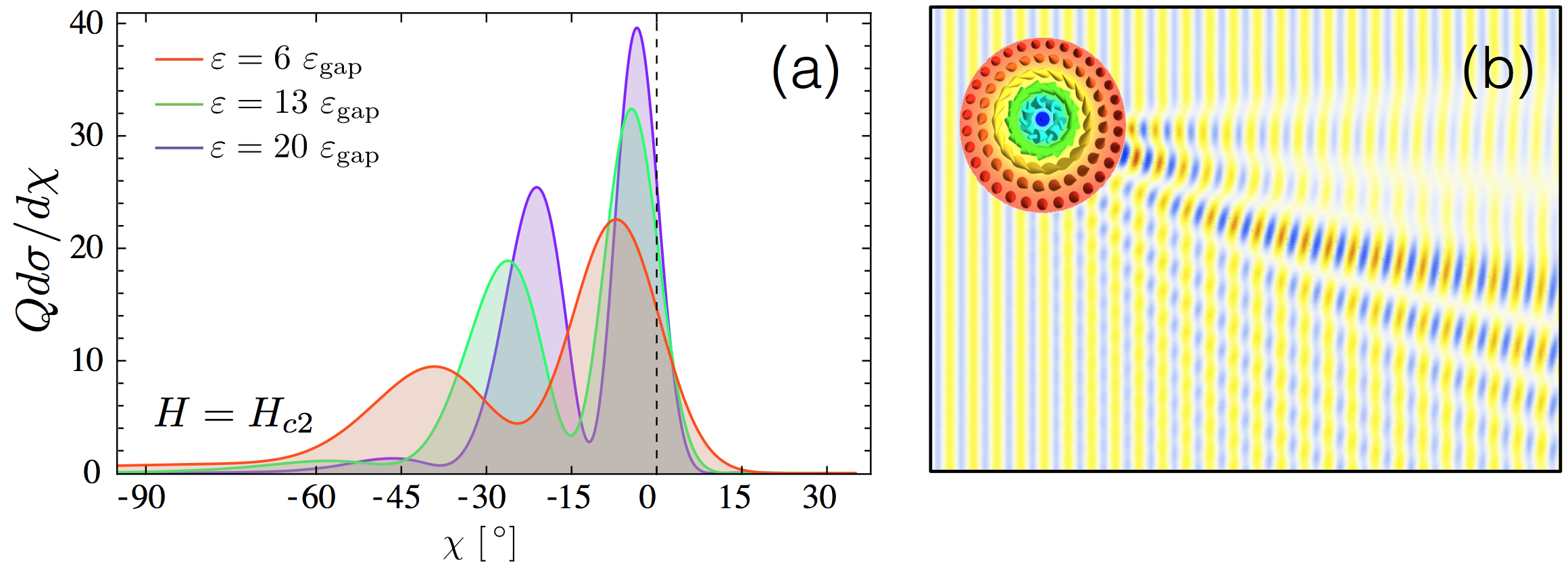}
\caption{(a) Scattering cross section of a skyrmion at $H=H_{c2}$ as a function of in-plane polar angle $\chi$ for a monochromatic spin wave with three different energies $\varepsilon$ \cite{Schuette2014}. It is characterized by skew scattering and oscillations. (b) Stationary magnon wavefunction with $\varepsilon = 20 \varepsilon_{\rm gap}$ in the presence of a skyrmion indicating that spin waves are preferentially scattered to the right-hand side of the skyrmion. Without the skyrmion the spin wave would propagate horizontally from left to right.
}
\label{Fig:SkyrmionCrossSection}
\end{figure}

The non-trivial topology of the skyrmion texture has a striking impact on the properties of the scattering states. Figure~\ref{Fig:SkyrmionCrossSection}(a) shows the scattering cross section for magnons propagating in-plane with an energy that is large compared to the magnon gap: it is characterized by oscillations  and strongly asymmetric with respect to forward scattering $\chi = 0$, i.e., there is skew scattering. This is also illustrated in panel (b) that shows a monochromatic magnon wave travelling from the left to the right hand side and scattering off a skyrmion. Both aspects can be transparently discussed by considering the magnon-skyrmion scattering problem in the high-energy limit \cite{Schroeter:2015}. In this limit, the magnon wave function $\psi$ of Eq.~(\ref{Expansion}) is simply governed by the Schr\"odinger equation
\begin{equation}
i \hbar \partial_t \psi = \Big[\frac{(- i\hbar \vec \nabla_\perp - \vec A(\vec r))^2}{2 m} + \varepsilon_{\rm gap} \Big] \psi
\end{equation}
where, on the level of Eq.~(\ref{F0}), the magnon gap $\varepsilon_{\rm gap}$ is proportional to the magnetic field, i.e., $\varepsilon_{\rm gap} = g \mu_B \mu_0 H$. The high-energy magnon propagating within the plane perpendicular to the field, $\vec \nabla_\perp = (\partial_x,\partial_y,0)$,  only scatters off a vector potential $\vec A = (A_x, A_y,0)$. The associated flux density $\mathcal{B}_z = \partial_x A_y - \partial_y A_x$ is finite so that the magnon scatters off an emergent orbital magnetic field which is localized to the skyrmion area. The total flux
\begin{equation} \label{TotalFlux}
\int dx dy\, \mathcal{B}_z = 4\pi\hbar
\end{equation}
is quantized and related to the topological skyrmion charge of the soliton. The magnon experiences an emergent Lorentz force due to $\mathcal{B}_z$ that deflects its trajectory leading to the skew scattering. A finite density of skyrmions are thus expected to give rise to a topological magnon Hall effect \cite{Iwasaki2014}. For a given skyrmion configuration
this Hall effect and the resulting spin accumulation might be used to read out a propagating spin wave signal.

The oscillations in the scattering cross section of Fig.~\ref{Fig:SkyrmionCrossSection} (a) arise from a phenomenon known as rainbow scattering. The emergent field $\mathcal{B}_z = \mathcal{B}_z(\rho)$ only depends on the distance $\rho$ to the skyrmion center. As a consequence, classical magnon trajectories with impact parameters $\pm b$, that pass the skyrmion at the same distance but either on its right- or left-hand side, will see the same flux and experience the same deflection angle. These trajectories interfere and lead to oscillations in the scattering cross section.

The emergent orbital field $\mathcal{B}_z$ for magnon excitations is characteristic for a topologically non-trivial magnetic texture. If the texture itself is moving with a finite velocity, an emergent electric field also arises according to Faraday's law \cite{Schulz:2012}. This emergent electrodynamics is one of the exciting aspects of skyrmion textures, see Ref.~\cite{Nagaosa:2013} for a review.

An individual skyrmion might be stabilized dynamically in a ferromagnetic thin film by using a current-biased nanocontact providing STT locally. On the one hand, such skyrmions could serve as scattering centers as discussed above. On the other hand, it was predicted that, in the presence of DMI, the breathing mode of the individual skyrmion can serve as an efficient and tuneable microwave signal generator \cite{Zhou2015STT}.

\subsection{Artificially tailored magnonic crystals based on skyrmions}
The eigenmodes of an individual skyrmion within confined geometries like a disc were discussed in a series of theoretical works \cite{JVKim2014,Beg2017,Guslienko2017,Mruczkiewicz2016}. In Ref.~\cite{PhysRevB.93.174429} Mruczkiewicz {\em et al.} investigated the spin-wave band structure 
arising from a periodic chain of nanodisks containing individual skyrmions. In Ref.~\cite{Ma2015}, Ma {\em et al.~}studied theoretically a long stripe in which a one-dimensional lattice of periodically ordered skyrmions were introduced dynamically via nanocontacts. In both cases, interfacial DMI and ultrathin films were considered. The skyrmions induced magnonic band structures with a dispersive character and collective properties \cite{PhysRevB.93.174429,Ma2015}. The stripe-based magnonic crystal has the advantage that both dipolarly and exchange-coupled modes can be controlled via the one-dimensional skyrmion lattice, while in chains of separated nanodisks only the dipolar coupling promotes propagating spin waves. Consequently, the work of Ma {\it et al.}~addressed spin waves propagating along the channel with wavelengths down to about 50 nm, thereby entering the exchange dominated regime with spin-wave eigenfrequencies of several ten GHz~\cite{Ma2015}. In Ref.~\cite{PhysRevB.93.174429} in contrast, the allowed minibands with dispersive spin waves were near the original eigenfrequencies of the gyrotropic and breathing modes below 1 GHz and around 13 GHz, respectively. Group velocities were found to be small and only a few 10 m/s. These values reflect the small coupling between the skyrmion-containing nanodisks. In Ref.~\cite{Ma2015}, velocities are found to be an order of magnitude larger, offering a better signal transmission in a magnonic device.

\subsection{Towards thin films with bulk DMI}
Helimagnets incorporating bulk DMI have already been prepared via thin-film deposition techniques on different substrates. Typically, growth- and strain-induced anisotropies are found to modify the magnetic phase diagram with respect to the bulk material. In addition thin films often consist of domains in which the crystallites exhibit a different chirality. Preliminary measurements on the spin-wave damping in thin films of the helimagnet FeGe deposited by two different groups have provided damping parameters $\alpha$ that differ by almost two orders of magnitude \cite{Beg2017,Zhang2016}. Here further research is needed to optimize chiral magnets in thin-film technology and exploit the full potential of bulk DMI.

Before we discuss interesting properties that theory predicts for thins films with bulk DMI we comment on the role of thin films that host skyrmions due to interfacial DMI. Spin dynamics in metallic magnetic multilayers exhibiting relevant interfacial DMI has been explored both theoretically and experimentally. Theoretically it has been shown that such layers form narrow domains walls that channel or scatter spin waves \cite{PhysRevLett.114.247206,Borys2015}. In the saturated magnetic state the interfacial DMI 
has been predicted to induce nonreciprocal spin-wave dispersion relations \cite{Cortes-Ortuno2013} that can be tailored via different multilayer compositions. Inelastic scattering of both electrons and light has been applied to such materials systems and showed spin-wave spectra exhibiting different eigenfrequencies for opposing wave-vector directions consistent with theoretical considerations \cite{PhysRevLett.104.137203,Nembach2015,KaiDi2015}. Interfacial DMI thus enriches a magnonics-related thin-film technology beyond skyrmions \cite{Lee.nanolett.5b02732}. For instance, nanomagnonic waveguides with unidirectional spin-wave propagation become possible \cite{C4RA07326F}. However, metallic magnetic layers with DMI are expected to exhibit a broad linewidth \cite{PhysRevB.82.014428}, i.e., an increased damping of spin waves. We note that the microscopic mechanism behind interfacial DMI lies in spin-orbit coupling. In metallic ferromagnets, spin-orbit coupling leads to spin-wave damping due to precession-induced intra- and interband excitations of conduction electrons and subsequent electron scattering \cite{Faehnle2011}. At the same time, metallic multilayers with non-collinear spin structures might experience enhanced spin-wave damping via spin pumping \cite{PhysRevLett.88.117601,Nembach2013,2015SaitohJAPL}. We suppose damping parameters $\alpha$ of metallic thin films hosting magnetic skyrmions to be four to five orders of magnitude worse compared to YIG which is the prototypical magnetic material in microwave technologies exploiting spin-precessional motion. When aiming at low spin-wave damping we suggest to focus on thin-film materials where an insulating magnet carries the spin waves. In insulating magnets, excitations of conduction electrons are not relevant and spin-wave damping is potentially small.

\subsubsection{Spin waves in thin films: field-polarized and magnetic helix state}
\label{sec:SpinWavesThinFilms}

Propagating spin waves have been explored in the field-polarized phase of a thin platelet of \cso\ that was extracted from a large single crystal using focused ion beam etching. When an in-plane field was parallel to the spin-wave wave vector a nonreciprocal spin-wave dispersion relation was extracted from the experimental spectra \cite{Seki2016PRB}. \cso\ thus offers nonreciprocity of both transmitted electromagnetic waves and spin waves. A bulk sample of noncentrosymmetric LiFe$_5$O$_8$ (space group P4$_1$32) with DMI was reported to exhibit a nonreciprocal spin-wave dispersion relation as well \cite{IguchiPRB2015}.

We now turn to the discussion of the spin-wave spectrum in a film of a cubic chiral magnet described by Eqs.~(\ref{F0}) and (\ref{Fdip}). The film is assumed to possess a thickness $d$ that is comparable to an intrinsic length scale of the system, in particular, the pitch length $2\pi/Q$ (Tab. \ref{Table1}). In the following we further assume an infinitely large film whose normal is oriented along the $z$-axis. For the thin film, one might take into account an additional magnetic anisotropy in the free energy functional
$F_{\rm aniso} = \int d\vec r \mathcal{F}_{\rm aniso}$ with
\begin{equation}
\mathcal{F}_{\rm aniso} = K \hat n_z^2
\end{equation}
and anisotropy constant $K$. It corresponds to an easy-plane anisotropy for $K > 0$ and an easy-axis anisotropy for $K<0$.

\subsection*{Large out-of-plane field: field-polarized state}

The situation in thin films is generally complicated due to  the boundary conditions of Eq.~(\ref{BC}) that the magnetization must obey at the surfaces of the film. It is still tractable for a magnetic field applied along the film normal, i.e., along the $z$-axis, $\vec H = H \hat z$. In this case the magnetization is fully polarized at large fields and $\hat n_{\rm eq} = \hat z$. As the film is infinite and translationally invariant, we can perform a partial Fourier transform and consider the magnon wavefunction $\vec \Psi_{\vec k_{\perp}, \omega}(z)$ as a function of in-plane momentum $\vec k_\perp = (k_x,k_y,0)$, frequency $\omega$, and the spatial $z$ coordinate.
The stationary wave equation can be cast into the form
\begin{equation} \label{EVPFilm}
\hbar \omega \tau^z \vec \Psi_{\vec k_{\perp}, \omega}(z) =
\mathcal{H}_{\rm loc} \vec\Psi_{\vec k_{\perp}, \omega}(z) + \int_{0}^d dz' \mathcal{K}(\vec k_\perp, z-z')\vec\Psi_{\vec k_{\perp}, \omega}(z')
\end{equation}
for a film located within the interval $z \in [0,d]$. The local part of the Hamiltonian reads
\begin{equation}
\mathcal{H}_{\rm loc} = \mathcal{D}
(-\mathds{1} \partial^2_z
- i 2 Q \tau^z \partial_z)  + (\mathcal{D} k_\perp^2 + g \mu_B \mu_0 H_{\rm int} - \frac{g\mu_B 2 K}{M_s}) \mathds{1}
\end{equation}
with the internal field $H_{\rm int} = H - M_s$ corresponding to a demagnetization factor $N_z = 1$. The nonlocal part derives from the dipolar interaction and its kernel is given by
\begin{equation}
\mathcal{K}(\vec k_\perp, z) = \frac{g\mu_B \mu_0 M_s}{4 |\vec k_\perp|} e^{-|\vec k_\perp| |z|}
\left(
\begin{array}{cc}
k_+ k_- &k_-^2 \\
k_+^2 & k_+ k_-
\end{array}
\right)
\end{equation}
in the same notation as in Eq.~(\ref{PolBulk}).
Importantly, on the two surfaces of the film, $z = 0$ and $z = d$, the wavefunction must obey the boundary conditions that follow from Eq.~(\ref{BC}). In the absence of surface pinning, they are given by
\begin{equation} \label{BCFilm}
(- i \partial_z + Q \tau^z)\vec \Psi_{\vec k_{\perp}, \omega}(z)\Big|_{\rm surface} = 0.
\end{equation}

For zero in-plane momentum $\vec k_\perp = 0$ the non-local part vanishes, $\mathcal{K}({\bf 0},z) = 0$, and the eigenvalue problem becomes local. The eigenfunctions obeying the above boundary conditions are then readily obtained
\begin{equation} \label{EigenfunctionsFilm}
\vec \Psi_{\sigma,p}(z) = \frac{1}{N_p} \cos\left(\frac{\pi p}{d} z\right)
e^{- i Q \tau^z z} \left(\begin{array}{c} 1+\sigma \\ 1-\sigma
\end{array}\right)
\end{equation}
with $\sigma = \pm 1$ and a normalization constant $N_p$. These eigenfunctions describe the so-called perpendicular standing spin-wave (PSSW) modes of the film  \cite{Demokritov2001} in the presence of DMI specified by the discrete quantum number $p = 0,1,2,3,...$. The corresponding eigenfrequencies are $\omega_{\pm,p} = \pm \omega_{p}$ with
\begin{equation} \label{EEFilmp}
\hbar \omega_{p} = \mathcal{D} \left(\frac{\pi p}{d}\right)^2
+ g \mu_B \mu_0 (H - H_{c2,z}).
\end{equation}
We introduced the critical field $H_{c2,z} = M_s + \frac{2 K}{\mu_0 M_s} + \frac{\mathcal{D} Q^2}{g \mu_B \mu_0}$ that is assumed to be positive.

The spectrum given by Eq. (\ref{EEFilmp}) differs from that of a field-polarized ferromagnetic film. The mode $p=0$ does not correspond to the uniform mode but carries a finite wavevector. For decreasing $H$, its gap vanishes at the critical field $H_{c2,z}$ triggering the formation of the helical state. The boundary conditions of Eq. (\ref{BCFilm}) result in the additional oscillating factor $e^{- i Q z}$ (for $\sigma = 1$). It was pointed out in Ref.~\cite{Zingsem} that the eigenfunction given by Eq. (\ref{EigenfunctionsFilm}) corresponds to the superposition of two non-reciprocal spin waves with wavevectors $\frac{\pi p}{d} + Q$ and $- \frac{\pi p}{d} + Q$. Importantly, the dynamic correction to the uniform magnetization is proportional to the integral $\int_0^d dz \vec \Psi_{+,p}(z)$ that is finite for all quantum numbers $p$ due to the additional phase factor $e^{- i Q z}$. As a result, all
PSSW modes couple to a uniform oscillating magnetic field \cite{Zingsem}. This is in contrast to a ferromagnetic film for which only the Kittel mode is excited by a uniform oscillating field if spins were free at the surfaces.

In the presence of a finite in-plane momentum $\vec k_\perp$ the nonlocal part $\mathcal{K}$ of the wave equation (\ref{EVPFilm}) must be considered.
This is usually done either approximately or numerically, and, for ferromagnets, it gives rise to the magnetostatic forward volume mode (MSFVM). Recent microwave absorption experiments on films of FeGe for an applied out-of-plane field have found two resonances at high fields that were interpreted as PSSW modes \cite{Zhang2016}.

\subsection*{Large in-plane field: twisted surface spins}

For a large magnetic field within the plane of the film, say, along the $x$-axis, the magnetization is not completely polarized and $\hat n_{\rm eq} \neq \hat x$. It is twisted close to the film surfaces due to the boundary conditions given by Eq. (\ref{BC}), and this boundary twist was experimentally investigated, e.g., in a film of MnSi \cite{Meynell2014} as well as in nanowires and lamellae of FeGe \cite{Du2015,Leonov2016}. The influence of this twist on the magnon spectrum was addressed theoretically in Refs.~\cite{Garcia-Sanchez2014,Mueller2016} neglecting however dipolar interactions. There it was found that the twist acts as an attractive potential for the exchange spin waves resulting in modes localized to the two surfaces. The mechanism for localization is thus different from surface- and edge confined magnetostatic modes already explored in magnonics, i.e., Damon-Eshbach modes \cite{Dam61} and modes in spin-wave wells \cite{Jorzick2002}, respectively.
Intriguingly, a recent theoretical work predicts that in the presence of a surface twist and corresponding edge magnons a specific field protocol can be implemented to create a chain of skyrmions at the edge of a field-polarized two-dimensional chiral magnet \cite{Mueller2016}.
Omitting the surface twist, Ref.~\cite{Cortes-Ortuno2013} theoretically discussed the full dipole-exchange spin-wave spectrum. A numerical micromagnetic simulation of the magnon spectrum in this configuration was recently presented in Ref.~\cite{Turgut2016} together with a comparison to microwave absorption experiments on FeGe films. Measurements  in this geometry have been also performed on Cu$_2$OSeO$_3$ \cite{Seki2016PRB}.

\subsection*{Thin film with spin-helix state}

We now turn to a discussion of spin-wave properties in the spin helix state. The treatment for a general orientation of the magnetic field is again complicated by the boundary conditions of Eq. (\ref{BC}). Only for a magnetic field aligned along the film normal ($z$-axis), the conical helix of Eq.~(\ref{ConHelix}) automatically fulfils the boundary conditions of Eq.~(\ref{BC}). We restrict our discussion to the case of vanishing in-plane magnon momentum $\vec k_\perp = 0$. Following similar steps as in the context of Eq.~(\ref{EEFilmp}), the eigenfrequencies of the PSSW modes are found to read
\begin{equation} \label{EEFilmh}
\hbar \omega_{p} = \sqrt{\mathcal{D}} \frac{\pi p}{d} \sqrt{\mathcal{D} \left(\frac{\pi p}{d}\right)^2 + g \mu_B \mu_0 H_{c2,z} \Big(1 - \Big(\frac{H}{H_{c2,z}}\Big)^2\Big)}.
\end{equation}
with $p = 0,1,2,...$ and the critical field $H_{c2,z} = M_s + \frac{2 K}{\mu_0 M_s} + \frac{\mathcal{D} Q^2}{g \mu_B \mu_0} > 0$. This formula is valid for $H \leq H_{c2,z}$. At the critical field $H_{c2,z}$ the eigenfrequencies of Eq. (\ref{EEFilmh}) directly connect to Eq.~(\ref{EEFilmp}) for each $p$ smoothly.

For a magnetic field within the plane of the film, different regimes might be realized.
For strong easy-plane anisotropy [\footnote{The effective anisotropy is determined here by the combination of the explicit anisotropy $K$ and the dipolar interaction, and it is quantified by $K_{\rm eff} = K + \mu_0 M_s^2/2$}], the helix axis might remain aligned along the film normal. Similar to monoaxial chiral magnets \cite{2012:Togawa:PhysRevLett}, the in-plane field then distorts the helix to a chiral soliton lattice, which has been recently observed
in films of FeGe \cite{Kanazawa2016}. The resonances of the chiral soliton lattice have been studied theoretically in Ref.~\cite{Kishine2009}.
In the other limit of  easy-axis anisotropy, the helix axis rotates into the film plane, and the boundary conditions (\ref{BC}) result in a deformation of the helix close to the surfaces. Even more complicated situations arise for weak or canted magnetic fields.

\normalsize
\section{Concluding remarks}
We reviewed the collective spin excitations in chiral magnets hosting the skyrmion lattice phase. We discussed how noncollinear spin structures such as spin helices and skyrmions enrich the dynamic properties of magnetic materials. We argued that beyond spintronics also magnonics is expected to benefit from developments in skyrmionics. One can anticipate for instance nonreciprocal microwave components, magnonic crystals and grating couplers controlled by magnetic and electric fields, nanoscale microwave generators, and Hall effect readout of spin-wave signals. Bulk chiral magnets with small damping open novel perspectives for the control and manipulation of both spin waves and electromagnetic waves in solids.
\section*{Acknowledgements}
M.G. acknowledges support from the Deutsche Forschungsgemeinschaft (DFG) via SFB 1143 ''Correlated Magnetism: From Frustration to Topology'' and grant GA 1072/5-1. D.G. acknowledges support by the DFG via the Transregio TRR80 ''From electronic correlations to functionality''. The Swiss National Science Foundation (SNSF) funds magnonics research on skyrmion-hosting materials via sinergia grant CRSII5-171003.
\appendix
\section{Dipolar matrix elements}
\label{appendix}

Here, we present the matrix elements $\mathcal{H}^{\alpha \beta}_{\rm dip}(\vec k)$ entering the spin wave equation of Eq.~(\ref{WEHelix}). We will use the matrix representation
\begin{equation}
\mathcal{H}_{\rm dip}(\vec k) = \left(\begin{array}{ccc}
\mathcal{H}_{\rm dip}^{1,1}(\vec k)	&	\mathcal{H}_{\rm dip}^{1,0}(\vec k)	&	\mathcal{H}_{\rm dip}^{1,-1}(\vec k)	\\
\mathcal{H}_{\rm dip}^{0,1}(\vec k)	&	\mathcal{H}_{\rm dip}^{0,0}(\vec k)	&	\mathcal{H}_{\rm dip}^{0,-1}(\vec k)	\\
\mathcal{H}_{\rm dip}^{-1,1}(\vec k)	&	\mathcal{H}_{\rm dip}^{-1,0}(\vec k)	&	\mathcal{H}_{\rm dip}^{-1,-1}(\vec k)	
\end{array}\right).
\end{equation}
For zero wavevectors, $\vec k = 0$, the matrix elements depend on the demagnetization factors $N_x$, $N_y$ and $N_z$,
\begin{equation}\label{eq:Vdem}
\hspace{-2.5cm} \mathcal{H}_{\rm dip}(\vec{0}) = \frac{\mathcal{D} Q^2 \chi_{\rm con}^{\rm int}}{2}\\
	{{
			\thinmuskip =0mu  	
			\medmuskip =1mu	
			\thickmuskip =2mu	
			\tiny
			\arraycolsep=-1pt\def\arraystretch{2.2}
			\left(\begin{array}{ccc}		
				\frac{N_x+N_y}{2} \left(\frac{1+c^2}{2}\mathds{1}+\frac{s^2}{2}\tau^x + c\,\tau^z \right) &
				0 &
				\frac{N_x-N_y}{2} \left( -\frac{s^2}{2}\mathds{1} - i c \,\tau^y - \frac{1+c^2}{2}\tau^x \right) \\
				0 &
				N_z s^2 (\mathds{1}-\tau^x) &
				0 \\
				\frac{N_x-N_y}{2} \left( -\frac{s^2}{2}\mathds{1} + i c\,\tau^y - \frac{1+c^2}{2}\tau^x \right) &
				0 &
				\frac{N_x+N_y}{2} \left( \frac{1+c^2}{2}\mathds{1} + \frac{s^2}{2}\tau^x - c\,\tau^z \right)
			\end{array}\right)
	}},
\end{equation}
where we abbreviated $s=\sin\theta$ and $c=\cos\theta$.

For wavevectors, $|\vec k| \gg 1/L$, large compared to the inverse linear size of the sample $L$, the matrix elements are given by
\begin{equation}\label{eq:Vdipdip2}
\hspace{-2.5cm}	
\mathcal{H}_{\rm dip}(\vec k) = \frac{\mathcal{D}Q^2 \chi_{\rm con}^{\rm int}}{4}
{{
		\thinmuskip =0mu  	
		\medmuskip =1mu	
		\thickmuskip =1mu	
		\tiny
		\arraycolsep=0pt\def\arraystretch{2.2}
		\left(\begin{array}{ccc}		
		k_- k_+ \left(\frac{1+c^2}{2}\mathds{1}+\frac{s^2}{2}\tau^x + c \,\tau^z \right) &
		k_- k_z \big( -c s(\mathds{1}-\tau^x) + i s (\tau^y +i\tau^z) \big) &
		k_- k_- \left( -\frac{s^2}{2}\mathds{1} - i c \,\tau^y - \frac{1+c^2}{2}\tau^x \right) \\
		k_z k_+ \big( -c s(\mathds{1}-\tau^x) - i s (\tau^y -i\tau^z) \big) &
		2\; \; k_z k_z s^2 (\mathds{1}-\tau^x) &
		k_z k_- \big( -c s(\mathds{1}-\tau^x) + i s (\tau^y -i\tau^z) \big) \\
		k_+ k_+ \left( -\frac{s^2}{2}\mathds{1} + i c \,\tau^y - \frac{1+c^2}{2}\tau^x \right) &
		k_+ k_z \big( -c s(\mathds{1}-\tau^x) - i s (\tau^y +i\tau^z) \big) &
		k_+ k_- \left( \frac{1+c^2}{2}\mathds{1} + \frac{s^2}{2}\tau^x - c\,\tau^z \right)
		\end{array}\right)
	}}
	\end{equation}
	where we used $k_{\pm} =  k_x \pm i k_y$.
 Note, that only the center element remains finite for $\vec k\parallel\hat z\parallel\vec Q$, particularly if $\vec k=n\vec Q$ for $n\in\mathds{Z}$ and $n\neq 0$.

\section*{References}
\bibliography{RefsSkxMagnonics}

\end{document}